\documentclass[twocolumn,pra,showpacs]{revtex4}

\usepackage{amssymb}
\usepackage{amsfonts}
\usepackage{amsmath}
\usepackage{graphicx}

\input{tcilatex}
\begin{document}

\title{Non-Markovian non-stationary completely positive open quantum system
dynamics}
\author{Adri\'{a}n A. Budini $^{1,2}$ and Paolo Grigolini $^{2,3,4}$}
\affiliation{$^{1}$Consejo Nacional de Investigationes Cient\'{\i}ficas y T\'{e}cnicas,
Centro At\'{o}mico Bariloche, Avenida E. Bustillo Km 9.5 (8400) Bariloche,
Argentina}
\affiliation{$^2$Center for Nonlinear Science, University of North Texas, P.O. Box
311427, Denton, Texas 76203-1427, USA}
\affiliation{$^3$Istituto dei Processi Chimico Fisici del CNR, Area della Ricerca di
Pisa, Via G. Moruzzi, 56124, Pisa, Italy}
\affiliation{$^{4}$Dipartimento di Fisica \textquotedblleft E.Fermi" - Universit\'{a} di
Pisa, Largo Pontecorvo, 3 56127, Pisa, Italy}
\date{\today}

\begin{abstract}
By modeling the interaction of a system with an environment through a
renewal approach, we demonstrate that completely positive non-Markovian
dynamics may develop some unexplored non-standard statistical properties.
The renewal approach is defined by a set of disruptive events, consisting in
the action of a completely positive superoperator over the system density
matrix. The random time intervals between events are described by an
arbitrary waiting-time distribution. We show that, in contrast to the
Markovian case, if one performs a system-preparation (measurement) at an
arbitrary time, the subsequent evolution of the density matrix evolution is
modified. The non-stationary character refers to the absence of an
asymptotic master equation even when the preparation is performed at
arbitrary long times. In spite of this property, we demonstrate that
operator expectation values and operators correlations have the same
dynamical structure, establishing the validity of a non-stationary quantum
regression hypothesis. The non-stationary property of the dynamic is also
analyzed through the response of the system to an external weak perturbation.
\end{abstract}

\maketitle

\section{Introduction}

\label{introduction}

The theory of Markovian open quantum systems \cite{Breuerbook} is well
established from both a mathematical and a physical point of view. The
theory of quantum dynamical semigroups, casting the structure of completely
positive (CP) trace-preserving maps, establishes that the
Kossakowski-Lindblad equations are the most general admissible forms of
evolution of the system density matrix. The application of these equations
ranges from quantum optics \cite{carmichael} to quantum information theory 
\cite{nielsen}.

As far as the quantum non-Markov case is concerned, there exist different
physical situations, and as a consequence a large variety of formalisms,
from which a solid proposal for an approach to non-Markovian quantum
dynamics \cite{Breuerbook,weiss} may emerge. A promising direction is
afforded by the non-Markovian generalization of the Kossakowski-Lindblad
equations. In a recent contribution \cite{barnett}, Barnett and Stenholm
showed that the adoption of a time convolution between a memory kernel and a
Kossakowski-Lindblad operator, although appealing, may lead to unphysical
results. However, their attempt attracted the attention of many researchers
to the search of the proper memory kernel for the time convoluted
Kossakowski-Lindblad equations \cite%
{wilkie,AClass,cresser,salo,lidar,maniscalco,kosa,semiMarkov,whitney,LindbladRate,JSP-QRT,breuer}%
. The main focus of most of these papers has been devoted to the search of
memory kernels that guarantees the CP condition of the solution map \cite%
{wilkie,AClass,cresser,salo,lidar,maniscalco,kosa,semiMarkov,whitney}. On
the other hand, different microscopic interactions that lead to the
convolution structure were established \cite{LindbladRate,JSP-QRT,breuer}
and applied in the characterization of spin environments \cite{breuerspin},
quantum Boltzman equations with internal degrees of freedom \cite{vacchini},
mesoscopic systems \cite{esposito}, as well as to fluorescent systems
coupled to complex self-fluctuating environments \cite{SingleMolecule}.

It is worth pointing out that the adoption of a renewal approach based on
the extension to quantum mechanics of the celebrated continuous time random
walk \cite{ctrw} leads naturally to the time convoluted structure that has
been originally hypothesized by the authors of Ref. \cite{barnett}, with no
risk of violating the CP condition \cite{AClass}. In fact, the result is
obtained through an average over infinitely many trajectories, each of which
consists in a series of sporadic and consecutive transformations
(system-environment collisions) of the system density matrix. The CP
condition is trivially satisfied when each collision is written in terms of
a CP transformation.

The continuous time random walk formalism has become a fundamental tool when
describing classical (non-Markovian) complex systems. In particular, the
existence of processes without a characteristic time scale (i.e.,
characterized by power law behaviors) has lead to an intensive review of the
formalism and of its possible extensions. One of the recent motivations for
studying that regime comes from the emergence of non-stationary phenomena in
the fluorescence intensity produced by (blinking) nanocrystal quantum dots
under laser radiation \cite%
{brokmann,GrigoAbsorption,verberkorrit,MarcusNature}. These experiments,
have led many researchers to revisit some basic tenets and tools of
equilibrium and non-equilibrium statistical mechanics, which are closely
related the one to the other: the Onsager principle \cite%
{onsager,CorrelationAge,agingCTRW}, non-stationary master equations \cite%
{agingCTRW}, linear response theory \cite%
{Kubo,KuboBook,GrigoLinearResponse,GrigoFluctDiss,EuroPhys,SokolovKlafer,error1,csf,Grigolini}%
, Wiener-Kinchine theorem \cite{WienerKinchine} and the ergodic hypothesis 
\cite{margolin}. The extension of these principles and theoretical tools for
dealing with non-stationary phenomena would be an outstanding breakthrough
in statistical mechanics. Besides the theoretical interest, there is an
increasing number of experimental situations \cite{Experiments} that can
draw benefits from that theoretical progress. While there exist different
issues that remain open, we are naturally challenged to find a proper
generalization to open quantum systems dynamics.

The main goal of this paper is to show that non-Markovian CP master
equations may fit non-standard non-stationary statistical phenomena, and
then to analyze the validity or extension of two cornerstones of the theory
of open quantum systems, i.e., the quantum regression hypothesis \cite%
{carmichael,lax} and linear response theory \cite{Kubo,KuboBook}. The
underlying dynamics of the system are defined by the renewal approach
introduced in Ref. \cite{AClass}.

To address the definition (in the context of an open quantum system theory)
of a non-stationary decay (or a non-stationary quantum master equation), we
introduce a system preparation at an arbitrary time posterior to the initial
coupling between the system and the bath. With preparation we means an
instantaneous CP operation (like a measurement or any sudden CP
transformation) that leaves the system in an arbitrary state. Then, two
times are introduced. One of them ($t$) \cite{AgingTime} measures the time
at which the preparation occurs, and the second one ($\tau $) measures the
time since the preparation. We shall use the term \textit{stationary decay}
(stationary master equation) to denote a relaxation after the preparation
done at time $t$ that is independent of $t,$ i.e., it only depends on $\tau
. $ On the same token, we shall use the term \textit{non-stationary decay}
(non-stationary master equation) to denote relaxation processes whose form
depends on $t,$ i.e., its functional dependence on $\tau $ is parametrized
by $t.$ In the Markovian case, the preparation always leads to the same
stationary master equation. In the non-Markovian case, we show that even
when the preparation is performed at arbitrary long times, the ensuing
relaxation may or not reach a stationary regime.

We also show that even in the presence of non-stationary effects, operators
expectation values and correlations have the same dynamical structure,
providing a generalization of the standard quantum regression theorem \cite%
{Breuerbook,carmichael} to a class of non-stationary quantum dynamics. The
response of the system to an external weak perturbation, while it can be
defined in terms of operator correlations \cite{Kubo,KuboBook}, generates
strong deviations with respect to the Markovian case.

The outline of this paper is as follows. In Sec. II, we review the renewal
approach and show how the non-stationary effects arise. In Sec. III we
obtain the evolution of both operator expectation values and correlations,
which allows us to establish a quantum regression theorem. In Sec. IV, the
response to external perturbations is studied. In Sec. V we provide the
conclusions.

\section{Non-stationary density matrix evolution}

\label{section2}

In the quantum application of the renewal approach \cite{AClass}, the
density matrix $\rho _{S}(\tau )$ of an open quantum system $S$ is
determined by means of an average over an ensemble of infinitely many
stochastic realizations, $\rho _{S}(\tau )=\langle \rho _{st}(\tau )\rangle
, $ where $\left\langle \cdots \right\rangle $ denotes the average over
realizations and $\rho _{st}(\tau )$ is the stochastic state associated to
each trajectory. They consist of a sequence of disruptive (collisional)
events occurring at random times. The times elapsed between two consecutive
events are randomly drawn from a waiting-time distribution density $w(t),$
satisfying $w(t)\geq 0,$ and $\int_{0}^{\infty }w(t)dt=1.$ Each event is
associated to an arbitrary CP transformation $\mathcal{E}$ of the system
state. It is defined by the Krauss form \cite{nielsen}%
\begin{equation}
\mathcal{E}[\rho ]=\sum_{i}C_{i}\rho C_{i}^{\dagger },  \label{SuperOperator}
\end{equation}%
where $\rho $\ is the system state prior to a given event. The operators $%
C_{i}$ satisfies the condition $\sum_{i}C_{i}^{\dagger }C_{i}=$I$.$
Furthermore, here we assume that between consecutive events, the evolution
of the system is defined by the propagator $\exp [t\mathcal{L}_{S}].$ The
superoperator $\mathcal{L}_{S}$ is the Liouville superoperator and
corresponds to a unitary transformation. Nevertheless, we remark that most
of the results hereby developed also apply when $\mathcal{L}_{S}$ is a
standard Lindblad superoperator, i.e., when the evolution between events
corresponds to a Markovian (CP) dissipative evolution.

By construction, each realization, and as a consequence the average over the
realizations, guarantees the CP condition of the solution map. The system's
dynamics begin at time $t=0$ (system-environment coupling). As stressed in
the introduction, we let the system evolve up to the time $t>0$ that we set
to be the new origin of time. The earlier work of Ref. \cite{AClass} is
confined to the condition $t=0$ and the main aim of this paper is to solve
the non-stationary issues raised by the condition $t>0.$ The main idea of
the method that we use is as follows. First of all we study the time
evolution of $\rho _{S}$ from $\rho _{S}(0)$ to $\rho _{S}(t+\tau ),$ and we
interpret the exact expression of $\rho _{S}(t+\tau )$ as the the density
matrix $\rho _{S}(\tau )$ that will be expressed in terms of the initial
condition $\rho (t).$ It is straightforward to get the exact expression of $%
\rho _{S}(\tau ),$ which reads%
\begin{equation}
\rho _{S}(\tau )=\sum_{n=0}^{\infty }\int_{0}^{\tau +t}dt^{\prime }\mathcal{P%
}_{0}(\tau +t-t^{\prime })\mathcal{W}^{(n)}(t^{\prime })\rho _{S}(0).
\label{Average}
\end{equation}%
The superoperator $\mathcal{W}^{(n)}(\tau )$ is defined in the Laplace
domain, $(\tau \rightarrow u,$ $t\rightarrow z)$ as%
\begin{equation}
\mathcal{W}^{(n)}(u)\equiv \lbrack \mathcal{E}w(u-\mathcal{L}_{S})]^{n},
\label{Woperator}
\end{equation}%
while the superoperator $\mathcal{P}_{0}(\tau )$ reads%
\begin{equation}
\mathcal{P}_{0}(u)\equiv P_{0}(u-\mathcal{L}_{S}),  \label{SurvivalOperator}
\end{equation}%
where $P_{0}(u)$ is the survival probability associated to $w(u),$ i.e.,%
\begin{equation}
P_{0}(u)\equiv \frac{1-w(u)}{u}.  \label{PoOperator}
\end{equation}%
The expression given by Eq.~(\ref{Average}) is a sum over all possible
realizations, each of them corresponding to a stochastic process with $n$
collisions. We have now to express it in terms of the initial condition%
\begin{equation}
\rho _{S}(t)=\sum_{n=0}^{\infty }\int_{0}^{t}dt^{\prime }\mathcal{P}%
_{0}(t-t^{\prime })\mathcal{W}^{(n)}(t^{\prime })\rho _{S}(0).
\label{InitialRenewal}
\end{equation}%
By using recursively the relation%
\begin{eqnarray}
\mathcal{W}^{(n)}(\tau +t) &=&\int_{0}^{\tau }d\tau ^{\prime }\mathcal{W}%
(\tau -\tau ^{\prime })\mathcal{W}^{(n-1)}(\tau ^{\prime }+t) \\
&&+\int_{0}^{t}dt^{\prime }\mathcal{W}(\tau +t-t^{\prime })\mathcal{W}%
^{(n-1)}(t^{\prime }),  \notag
\end{eqnarray}%
we rewrite Eq.~(\ref{Average}) as%
\begin{eqnarray}
\rho _{S}(\tau ) &=&\Xi _{0}(\tau ,t)+\sum_{n=1}^{\infty }\int_{0}^{\tau
}d\tau ^{\prime }\int_{0}^{\tau ^{\prime }}d\tau ^{\prime \prime }
\label{Rho(tau)} \\
&&\times \mathcal{P}_{0}(\tau -\tau ^{\prime })\mathcal{W}^{(n-1)}(\tau
^{\prime }-\tau ^{\prime \prime })\Xi _{w}(\tau ^{\prime \prime },t),  \notag
\end{eqnarray}%
where we have defined%
\begin{eqnarray}
\Xi _{0}(\tau ,t) &=&\sum_{n=0}^{\infty }\int_{0}^{t}dt^{\prime }\mathcal{P}%
_{0}(\tau +t-t^{\prime })\mathcal{W}^{(n)}(t^{\prime })\rho _{S}(0),\ \ \ \ 
\\
\Xi _{w}(\tau ,t) &=&\sum_{n=0}^{\infty }\int_{0}^{t}dt^{\prime }\mathcal{W}%
(\tau +t-t^{\prime })\mathcal{W}^{(n)}(t^{\prime })\rho _{S}(0).\ \ \ \ 
\end{eqnarray}%
The Laplace transform of $\rho _{S}(\tau )$ of Eq.~(\ref{Rho(tau)}) then
reads%
\begin{equation}
\rho _{S}(u)=\Xi _{0}(u,z)+\sum_{n=1}^{\infty }\mathcal{P}_{0}(u)\mathcal{W}%
^{(n-1)}(u)\Xi _{w}(u,z).  \label{LaplaceRenewal}
\end{equation}%
By using the relation%
\begin{equation}
\int_{0}^{\infty }d\tau \int_{0}^{\infty }dte^{-u\tau }e^{-zt}g(\tau +t)=%
\frac{g(u)-g(z)}{z-u},  \label{DoubleLaplace}
\end{equation}%
which is valid for any arbitrary function $g(t)$, with the notation $%
g(u)\equiv \int_{0}^{\infty }d\tau e^{-u\tau }g(\tau )$ and $g(z)\equiv
\int_{0}^{\infty }dte^{-zt}g(t)$, the Laplace transforms of $\Xi _{0}(\tau
,t)$ and $\Xi _{w}(\tau ,t)$ are written as%
\begin{eqnarray}
\Xi _{0}(u,z) &=&\frac{\mathcal{P}_{0}(u)-\mathcal{P}_{0}(z)}{z-u}\frac{1}{%
\mathcal{P}_{0}(z)}\rho _{S}(z), \\
\Xi _{w}(u,z) &=&\frac{\mathcal{W}(u)-\mathcal{W}(z)}{z-u}\frac{1}{\mathcal{P%
}_{0}(z)}\rho _{S}(z).
\end{eqnarray}%
Here, $\rho _{S}(z)$ is the Laplace transform of $\rho _{S}(t),$ given by
Eq.~(\ref{InitialRenewal}). By plugging these expressions into Eq.~(\ref%
{LaplaceRenewal}), after some algebra we get%
\begin{equation}
\rho _{S}(u)=\mathcal{G}(u)[\rho _{S}(z)+\mathrm{I}_{\rho }(u,z)],
\label{RhoLaplace}
\end{equation}%
where the propagator $\mathcal{G}(u)$ is defined by%
\begin{equation}
\mathcal{G}(u)=\frac{1}{u-[\mathcal{L}_{S}+\mathcal{L}K(u-\mathcal{L}_{S})]},
\label{PropaDensityMatrix}
\end{equation}%
and the inhomogeneous term by 
\begin{equation}
\mathrm{I}_{\rho }(u,z)=\mathcal{L}\Delta (u-\mathcal{L}_{S},z-\mathcal{L}%
_{S})(z-\mathcal{L}_{S})\rho _{S}(z).  \label{Inhomogeneous}
\end{equation}%
In the time domain, Eq.~(\ref{RhoLaplace}) becomes 
\begin{eqnarray}
\frac{d\rho _{S}(\tau )}{d\tau } &=&\mathcal{L}_{S}\rho _{S}(\tau
)+\int_{0}^{\tau }\!\!d\tau ^{\prime }K(\tau -\tau ^{\prime })\mathcal{L}%
e^{(\tau -\tau ^{\prime })\mathcal{L}_{S}}\rho _{S}(\tau ^{\prime })  \notag
\\
&&+\mathrm{I}_{\rho }(\tau ,t).  \label{Master}
\end{eqnarray}%
This equation is one of the central results of this section. It defines the
evolution of the average density matrix of the system in the interval $%
(t,t+\tau ),$ with the initial condition Eq.~(\ref{InitialRenewal}). The
superoperator $\mathcal{L}$ is defined by%
\begin{equation}
\mathcal{L}=\mathcal{E}-1,  \label{Lindblad}
\end{equation}%
which in turn can be written with the Lindblad structure%
\begin{equation}
\mathcal{L}[\bullet ]=\frac{1}{2}\sum_{i}\{[C_{i},\bullet C_{i}^{\dagger
}]+[C_{i}\bullet ,C_{i}^{\dagger }]\}.  \label{LindbladBaseC}
\end{equation}%
The memory kernel function $K(\tau )$ is defined in the Laplace domain by%
\begin{equation}
K(u)=\frac{uw(u)}{1-w(u)}.  \label{Kernel}
\end{equation}%
The inhomogeneous contribution $\mathrm{I}_{\rho }(u,z)$\ [Eq.~(\ref%
{Inhomogeneous})] is proportional to the function%
\begin{equation}
\Delta (u,z)=\frac{\tilde{w}(u,z)}{1-w(u)}-\frac{w(u)/z}{1-w(u)},
\label{Delta}
\end{equation}%
where the function $\tilde{w}(u,z)$\ reads 
\begin{equation}
\tilde{w}(u,z)=\frac{w(u)-w(z)}{z-u}\frac{1}{1-w(z)}.  \label{w(u,z)}
\end{equation}%
By using the relation of Eq.~(\ref{DoubleLaplace}), $\tilde{w}(u,z)$ can be
written in the time domain as%
\begin{equation}
\tilde{w}(\tau ,t)=w(\tau +t)+\sum_{n=1}^{\infty }\int_{0}^{t}dt^{\prime
}w(\tau +t-t^{\prime })w^{(n)}(t^{\prime }),  \label{AgedWait}
\end{equation}%
where $w^{(n)}(z)\equiv \lbrack w(z)]^{n}.$ This expression allows us to
interpret $\tilde{w}(\tau ,t)$ as a conditional waiting-time distribution
density, and more precisely as the probability distribution density of
meeting the first event at time $\tau ,$ given that the observation time (of
events) begins at time $t.$ The second term on the right hand side of Eq.~(%
\ref{AgedWait}) takes into account all possible events at times earlier than 
$t.$ Consistently, notice that for $t=0,$ $\tilde{w}(\tau ,0)=w(\tau ).$

In the time domain, the function $\Delta (u,z)$ read%
\begin{equation}
\Delta (\tau ,t)=f(\tau ,t)-f(\tau ,0),  \label{DeltainTime}
\end{equation}%
where we have introduced the (sprinkling) distribution 
\begin{equation}
f(u,z)=\frac{\tilde{w}(u,z)}{1-w(u)}.  \label{SprinklingLaplace}
\end{equation}%
By writing this expression in the time domain, 
\begin{equation}
f(\tau ,t)=\tilde{w}(\tau ,t)+\sum_{n=1}^{\infty }\int_{0}^{\tau }d\tau
^{\prime }w^{(n)}(\tau -\tau ^{\prime })\tilde{w}(\tau ^{\prime },t),
\label{Sprinkling}
\end{equation}%
it follows that $f(\tau ,t)d\tau $ is the probability of an event occurrence
in the time interval $(\tau ,\tau +d\tau ),$ given that the observation time
begins at time $t,$ regardless of whether or not any event occurred at
earlier times. It satisfies the relation $f(\tau ,0)=\tilde{w}(0,\tau ).$ On
the other hand, notice that the function $\Upsilon (u,z)\equiv z\Delta (u,z)$
[appearing in Eq.~(\ref{Inhomogeneous})] can be written in the time domain
as $\Upsilon (\tau ,t)=(d/dt)f(\tau ,t).$

Both the kernel $K(\tau )$ and the inhomogeneous contribution $\mathrm{I}%
_{\rho }(\tau ,t)$ are clear signatures of the non-Markovian property of the
evolution Eq.~(\ref{Master}). Using Eq.~(\ref{RhoLaplace}), it is easy to
realize that the evolution of $\rho _{S}(\tau )$ can always be rewritten as
an homogeneous evolution [see for example Eq.~(\ref{Homogenea})].
Nevertheless, the inhomogeneous structure allows us to understand which is
the effect of shifting the initial time condition from $\rho _{S}(0)$ to $%
\rho _{S}(t).$ In fact, Eq.~(\ref{Inhomogeneous}) tells us that in the
interval $(t,t+\tau )$ the departure of the system time evolution from its
time evolution in $(0,t)$ is measured by Eq.~(\ref{DeltainTime}).
Consistently, for $t=0,$ the inhomogeneous contribution vanishes, i.e., $%
\mathrm{I}_{\rho }(\tau ,0)=0.$

No departure of the system evolution in $(t,t+\tau )$ from the time
evolution in $(0,t)$ must occur in the Markovian case. This case is
recovered by assuming an exponential waiting-time distribution (Poisson
case), $w(\tau )=\gamma \exp [-\gamma \tau ].$ From Eqs.~(\ref{w(u,z)}) and (%
\ref{SprinklingLaplace}) it follows that $\tilde{w}(\tau ,t)=w(\tau )$\ and $%
f(\tau ,t)=\gamma ,$ thereby implying the vanishing of the inhomogeneous
term, and the relation $K(\tau )=\gamma \delta (\tau ),$ which turns Eq.~(%
\ref{Master}) into a standard Lindblad equation%
\begin{equation}
\frac{d\rho _{S}(\tau )}{d\tau }=[\mathcal{L}_{S}+\gamma \mathcal{L}]\rho
_{S}(\tau ).  \label{Markov}
\end{equation}%
On the other hand, we remark that in Eq.~(\ref{Master}), and in Eq.~(\ref%
{Inhomogeneous}) as well, the superoperator $\mathcal{L}$ may be assigned an
arbitrary Lindblad form. In fact, when $\mathcal{L}\neq \mathcal{E}-$\textrm{%
I} [see Eq.~(\ref{Lindblad})] the superoperator $\mathcal{E}$ [Eq.~(\ref%
{SuperOperator})] can be defined as $\mathcal{E}[\rho ]=\{\mathrm{I}%
+[e^{\kappa \mathcal{L}}-\mathrm{I}]\}\rho ,$ where $\kappa $ must be
interpreted as a control parameter. Then, we recover Eq.~(\ref{Master}) with
an arbitrary $\mathcal{L}$ in the limiting condition in which simultaneously 
$\kappa \rightarrow 0$ and the number of events per unit of time go to
infinity, the last limit being controlled by the distribution $f(\tau ,0)$
of Eq.~(\ref{Sprinkling}).

\subsection{Initial preparation at time t}

\label{newinitial}

The initial condition associated to Eq.~(\ref{Master}) is given by $\rho
_{S}(t),$ Eq.~(\ref{InitialRenewal}), which in turn carries information
about the system dynamics in the interval $(0,t).$ Therefore, Eq.~(\ref%
{Master}) does not give more information than a master equation describing
the evolution in the interval $(0,t+\tau ).$ Nevertheless, the master
equation (\ref{Master}) may acquire a different status if the initial
condition at time $t$ can be chosen as \textit{any} non-equilibrium form of
the density matrix $\rho _{S}.$ This is done by introducing the main
ingredient of our formalism, i.e., by adopting the concept of \textit{%
preparation}, namely, a change $\rho _{S}(t)\overset{\Pi }{\rightarrow }\rho
_{\Pi },\ $compatible with a CP transformation $\Pi .$ The role of the
preparation is to erase the dependence of the evolution on the previous
history of the system without erasing the memory of the universe, i.e., the
system-environment arrangement.

In the Laplace domain the preparation is defined by%
\begin{equation}
(z-\mathcal{L}_{S})\rho _{S}(z)\rightarrow \rho _{\Pi },  \label{Preparation}
\end{equation}%
which in the time domain yields $\rho _{S}(t)\rightarrow \rho _{\Pi }\exp [%
\mathcal{L}_{S}t].$ The extra unitary contribution is introduced to take
into account that the Hamiltonian evolution defined by $\mathcal{L}_{S}$
begins at time $t=0.$ Similarly, one can interpret Eq.~(\ref{Preparation})
as a preparation in an interaction representation with respect to $\mathcal{L%
}_{S}.$

Under the preparation condition of Eq.~(\ref{Preparation}), the time
evolution structure of Eq.~(\ref{Master}) is still valid, provided that the
initial condition is fixed to be $\rho _{\Pi }\exp [\mathcal{L}_{S}t],$ with
the inhomogeneous term now reading%
\begin{equation}
\mathrm{I}_{\rho }(\tau ,t)=\mathcal{L}\Delta (\tau ,t)\exp [(\tau +t)%
\mathcal{L}_{S}]\rho _{\Pi }.  \label{InhoPrepa}
\end{equation}%
We remark that the time evolution of Eq.~(\ref{Master}) with the
contribution of Eq.~(\ref{InhoPrepa}) remains a CP structure. In fact, also
its solution admits an interpretation in terms of trajectories that preserve
the CP condition.

The time evolution of the density matrix generated by Eq.~(\ref{Master}),
with the inhomogeneous contribution of Eq.~(\ref{InhoPrepa}), in principle
depends on the preparation time $t.$ Which is the form of the dependence of
the density time evolution in $(t,t+\tau )$ on $t?$ Of particular interest
is to assess under which conditions this dependence is lost, so as to
generate in the long-time limit the stationary behavior defined in Sec. \ref%
{introduction}. In the case where the solution becomes asymptotically
stationary, it is of interest to assess if this stationary time evolution ($%
t\rightarrow \infty $) is characterized by non-Markovian effects stronger or
weaker than the time evolution with the preparation stage coinciding with
the initialization stage, \textit{i.e.},with $t=0.$ These important
questions will be answered with the help of the simple examples discussed in
Section \ref{examples}.

As a last but not least remark of this Section, let us notice that the time
evolution of Eq.~(\ref{Master}), with the preparation condition of Eq.~(\ref%
{Preparation}), can be easily written in an equivalent form, as an
homogeneous time evolution, as follows%
\begin{eqnarray}
\frac{d\rho _{S}(\tau )}{d\tau } &=&\mathcal{L}_{S}\rho _{S}(\tau
)+\int_{0}^{\tau }d\tau ^{\prime }\int_{0}^{\tau ^{\prime }}d\tau ^{\prime
\prime }\mathcal{M}_{t}(\tau -\tau ^{\prime })  \notag \\
&&K_{t}(\tau ^{\prime }-\tau ^{\prime \prime })\mathcal{L}e^{(\tau ^{\prime
}-\tau ^{\prime \prime })\mathcal{L}_{S}}\rho _{S}(\tau ^{\prime \prime }),
\label{Homogenea}
\end{eqnarray}%
where $\mathcal{M}_{z}(u)=[1+\mathcal{L}\Delta (u-\mathcal{L}_{S},z-\mathcal{%
L}_{S})]^{-1},$ and $K_{t}(u)=u\tilde{w}(u,t)/[1-w(u)].$ While this
expression avoids the complication arising from the presence of an
inhomogeneous term, the kernel structure is more complicated, involving all
powers of the Lindblad superoperator $\mathcal{L}.$ Eq.~(\ref{Homogenea})
recovers and generalize the classical master equation obtained in Ref.~\cite%
{agingCTRW}.

\subsection{Examples}

\label{examples}

To make more transparent the spirit of the renewal approach of this paper,
here we illustrate it in action on two exemplary cases of the same simple
model. While the microscopic origin the superoperator $\mathcal{E}$ and the
waiting time distribution $w(t)$ is not completely understood \cite%
{AClass,LindbladRate}, from Eq.~(\ref{Master}) it becomes clear that the
former object defines the underlying Lindblad like structure Eq.~(\ref%
{LindbladBaseC}). Then, it establishes the coupling between the density
matrix elements. On the other hand, $w(t)$ can be settled in a
phenomenological way as a function of the characteristic system decay
behavior (see next examples). It leading property is the average waiting
time, $\int_{0}^{\infty }tw(t)dt,$ which may be finite or divergent, the
last case giving rise to strong non-stationary effects.

As a simple model, we consider a degenerate two-level system $(\mathcal{L}%
_{S}\rightarrow 0)$ and the superoperator%
\begin{equation}
\mathcal{E}[\bullet ]=\sigma _{z}\bullet \sigma _{z}.
\end{equation}%
The time evolution of the expectation values of the Pauli matrixes $\sigma
_{i},$ $S_{i}(\tau )\equiv \mathrm{Tr}_{S}[\rho _{S}(\tau )\sigma _{i}],$
with $i=x,y,z,$ is given by Eqs.~(\ref{Master}) and (\ref{InhoPrepa}) [or
equivalently by Eq.~(\ref{Homogenea})], and it reads%
\begin{equation}
\frac{dS_{X,Y}(\tau )}{d\tau }=-\int_{0}^{\tau }d\tau ^{\prime }\tilde{K}%
_{t}(\tau -\tau ^{\prime })S_{X,Y}(\tau ^{\prime }),  \label{Pauli}
\end{equation}%
while $S_{Z}(\tau )=S_{Z}(0).$ With $S_{i}(0)$ we denote the expectation
values after the preparation. The kernel is defined by its Laplace transform 
$\tilde{K}_{t}(u)=u\tilde{w}(u,t)/[1-\tilde{w}(u,t)].$ The solution of Eq.~(%
\ref{Pauli}) is%
\begin{equation}
S_{X,Y}(\tau )=S_{X,Y}(0)\tilde{P}_{0}(\tau ,t),  \label{Expectation}
\end{equation}%
where $\tilde{P}_{0}(\tau ,t)$ is the survival probability associated to $%
\tilde{w}(\tau ,t),$ i.e., $\tilde{P}_{0}(\tau ,t)=1-\int_{0}^{\tau }d\tau
^{\prime }\tilde{w}(\tau ^{\prime },t).$ It can be rewritten as%
\begin{equation}
\tilde{P}_{0}(\tau ,t)=P_{0}(\tau +t)+\int_{0}^{t}dt^{\prime }P_{0}(\tau
+t-t^{\prime })f(t^{\prime },0).  \label{SolutionWithSurvival}
\end{equation}%
Here, $P_{0}(\tau )$ is defined by its Laplace transform Eq.~(\ref%
{PoOperator}), while $f(t,0)$ follows from Eq.~(\ref{Sprinkling}), i.e., $%
f(z,0)=w(z)/[1-w(z)].$

The decay of the expectation values of Eq.~(\ref{Expectation}) depends on
both $\tau $\ and $t.$ Its explicit analytical form depends on the choice
done for the waiting-time distribution $w(t).$ As first case, we select the
bi-exponential case%
\begin{equation}
w(t)=P_{a}\gamma _{a}e^{-\gamma _{a}t}+P_{b}\gamma _{b}e^{-\gamma _{b}t},
\label{WaitBiexpo}
\end{equation}%
with $P_{a}+P_{b}=1.$ The distribution Eq.~(\ref{Sprinkling}) $(t=0)$ reads%
\begin{equation}
f(\tau ,0)=\left\langle \gamma \right\rangle \theta (\tau )-[\left\langle
\gamma \right\rangle -\left\langle \tau \right\rangle ^{-1}](1-e^{-\eta \tau
}),  \label{RocioTwoExpo}
\end{equation}%
where $\theta (\tau )$ is the step function and\ we have introduced the
parameters $\left\langle \gamma \right\rangle \equiv P_{a}\gamma
_{a}+P_{b}\gamma _{b},$ $\left\langle \tau \right\rangle \equiv P_{a}\gamma
_{a}^{-1}+P_{b}\gamma _{b}^{-1}=\int_{0}^{\infty }\tau w(\tau )d\tau <\infty
,$ and $\eta \equiv P_{a}\gamma _{b}+P_{b}\gamma _{a}.$ Notice that after a
transient of order $1/\eta ,$ the sprinkling distribution, as in the
Markovian case, is constant, i.e., $f(\tau ,0)\simeq 1/\left\langle \tau
\right\rangle >0.$

The coherence decay, independently of the time $t,$ can be written as%
\begin{equation}
\tilde{P}_{0}(\tau ,t)=P_{a}(t)e^{-\gamma _{a}\tau }+P_{b}(t)e^{-\gamma
_{b}\tau }.
\end{equation}%
All the dependence on the preparation time is carried out by the weights $%
P_{a}(t)$ and $P_{b}(t).$ Their explicit form follows straightforwardly from
Eq.~(\ref{SolutionWithSurvival}) as a superposition of exponential
functions. They satisfy the boundary conditions $P_{a}(0)=P_{a},$ and $%
P_{b}(0)=P_{b}.$ In the limit $t\rightarrow \infty ,$ the asymptotic
stationary decay reads%
\begin{equation}
\tilde{P}_{0}(\tau ,\infty )=\frac{P_{a}}{\left\langle \tau \right\rangle
\gamma _{a}}e^{-\gamma _{a}\tau }+\frac{P_{b}}{\left\langle \tau
\right\rangle \gamma _{b}}e^{-\gamma _{b}\tau }.
\end{equation}%
%
%
%
%
%
%
%
%
%
%
%
%figura%figura%figura%figura%figurav%figura%figura%figura%figura%figura%figura%figura%figura%figura%figurav%figura%figura%figura%figura%figura
%figura%figura%figura%figura%figurav%figura%figura%figura%figura%figura%figura%figura%figura%figura%figurav%figura%figura%figura%figura%figura
\begin{figure}[tb]
\includegraphics[bb=32 0 880 1210,angle=0,width=7.25 cm]{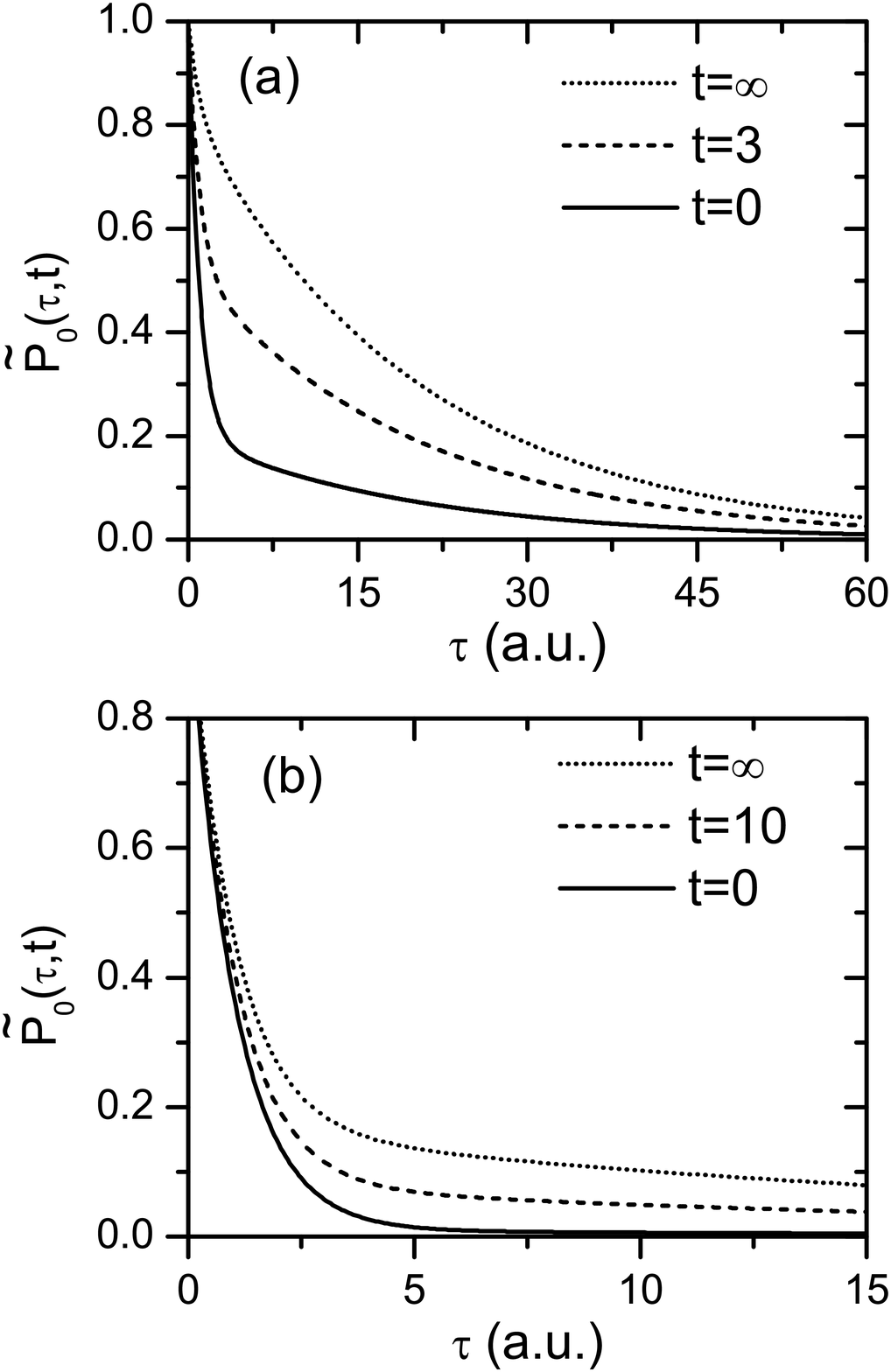}
\caption{Coherence decay $\tilde{P}_{0}(\protect\tau ,t)$ [Eq.~(\protect\ref%
{Expectation})], for different times $t,$ for the waiting time distribution
Eq.~(\protect\ref{WaitBiexpo}). In (a) the parameters are $P_{a}=0.8,$ $%
P_{b}=0.2.$ In (b) are $P_{a}=0.99,$ $P_{b}=10^{-3}.$ In both cases the
rates are $\protect\gamma _{a}=1,$ $\protect\gamma _{b}=0.05.$ Both, $%
\protect\tau $ and $t$\ are measured in arbitrary units (a.u.).}
\end{figure}
%figura%figura%figura%figura%figurav%figura%figura%figura%figura%figura%figura%figura%figura%figura%figurav%figura%figura%figura%figura%figura
%figura%figura%figura%figura%figurav%figura%figura%figura%figura%figura%figura%figura%figura%figura%figurav%figura%figura%figura%figura%figura
In Fig.~1, we show the decay defined by $\tilde{P}_{0}(\tau ,t)$ for
different preparation times $t,$ and for two different sets of
characteristic parameter values. After a transient of order $\eta ,$ both
cases reach a stationary decay regime. By comparing these figures one with
the other, we realize that the asymptotic decay may yield arbitrary
departures from the dynamics generated by setting $t=0.$ In fact, in
Fig.~1a, the asymptotic decay is almost exponential while the initial one is
bi-exponential. In Fig.~1b the inverse situation is observed. This simple
example demonstrates that no general conclusion can be drawn about the
properties of the stationary time evolution. 
%figura%figura%figura%figura%figurav%figura%figura%figura%figura%figura%figura%figura%figura%figura%figurav%figura%figura%figura%figura%figura
%figura%figura%figura%figura%figurav%figura%figura%figura%figura%figura%figura%figura%figura%figura%figurav%figura%figura%figura%figura%figura
\begin{figure}[tb]
\includegraphics[bb=20 4 828 583,angle=0,width=7.75 cm]{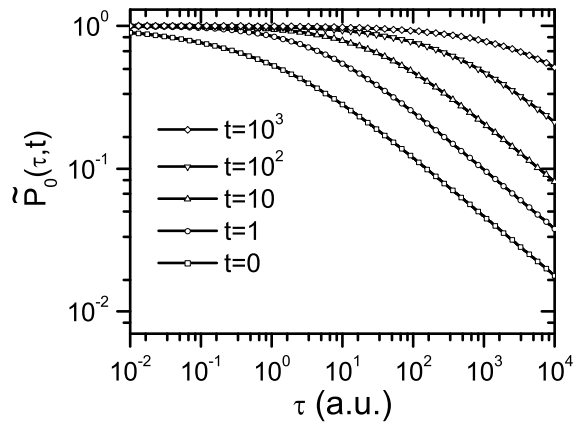}
\caption{Coherence decay $\tilde{P}_{0}(\protect\tau ,t)$ [Eq.~(\protect\ref%
{Expectation})], for different times $t,$ for the fractional waiting time
distribution Eq.~(\protect\ref{WaitFrac}). The parameters are $A_{\protect%
\alpha }=1,$ $\protect\alpha =1/2.$ Both, $\protect\tau $ and $t$\ are
measured in arbitrary units (a.u.).}
\end{figure}
%figura%figura%figura%figura%figurav%figura%figura%figura%figura%figura%figura%figura%figura%figura%figurav%figura%figura%figura%figura%figura
%figura%figura%figura%figura%figurav%figura%figura%figura%figura%figura%figura%figura%figura%figura%figurav%figura%figura%figura%figura%figura

As a second exemplary case, we consider the waiting-time distribution%
\begin{equation}
w(u)=\frac{A_{\alpha }}{A_{\alpha }+u^{\alpha }}.  \label{WaitFrac}
\end{equation}%
where the units of $A_{\alpha }$ are $1/sec^{\alpha },$ and $0<\alpha \leq
1. $ Note that for $\alpha =1$ this expression reduces to the Laplace
transform of an exponential function. The kernel Eq.~(\ref{Kernel}) read $%
K\left( u\right) =A_{\alpha }u^{1-\alpha }.$ As is well known \cite{metzler}%
, this kind of kernel is related to a fractional derivative operator. In
contrast to Eq.~(\ref{RocioTwoExpo}), here we get%
\begin{equation}
f(\tau ,0)=\frac{A_{\alpha }}{\Gamma (\alpha )}\frac{1}{\tau ^{(1-\alpha )}},
\label{SprinklingFraccionaria}
\end{equation}%
where $\Gamma (x)$\ is the Gamma function. Then, in this case, $\lim_{\tau
\rightarrow \infty }f(\tau ,0)=0.$ This property is directly related to the
divergence of the average period between events, i.e., $\int_{0}^{\infty
}\tau w(\tau )d\tau =\infty $ \cite{metzler}.

By using Eq.~(\ref{SolutionWithSurvival}) and the fact that $P_{0}(\tau )$
is, in this case, a Mittag-Leffler function \cite{AClass,metzler}, we can
write $\tilde{P}_{0}(\tau ,t)$ \cite{exact} under the form of a series
expansion. By using the property that for $A_{\alpha }\tau ^{\alpha }\gg 1,$ 
$P_{0}(\tau )\approx A_{\alpha }/[\tau ^{\alpha }\Gamma (1-\alpha )],$ when $%
\tau \gg t$ we get the following asymptotic expression%
\begin{equation}
\tilde{P}_{0}(\tau ,t)\!\approx \!\frac{1}{\Gamma (1-\alpha )}\left[ \frac{%
A_{\alpha }^{-1}}{(\tau +t)^{\alpha }}+\frac{1}{\alpha \Gamma (\alpha )}%
\frac{t^{\alpha }}{(\tau +t)^{\alpha }}\right] .  \label{Ajuste}
\end{equation}%
Therefore, in this case there not exists an asymptotic stationary decay. In
fact, this expression shows that at any time the decay dynamics depends on
the preparation time $t.$ In Fig.~2 we plot the function $\tilde{P}_{0}(\tau
,t),$ Eq.~(\ref{SolutionWithSurvival}), for different values of $t.$ Eq.~(%
\ref{Ajuste}) correctly fits their asymptotic decay behavior. Consistently,
we found that there not exists a stationary decay behavior.

\section{Regression hypothesis}

The generalization of the classical regression hypothesis\ \cite{onsager} to
a quantum context is called quantum regression theorem \cite{carmichael,lax}%
. It states that operator expectation values and operator correlations have
the same dynamical behavior. Here, we explore the possibility of
generalizing this theorem to the renewal case.

\subsection{Operators dual evolution}

In order to define operator correlations, we have to move from the Schr\"{o}%
dinger to the dual or Heisenberg representation. In the renewal case here
under study this corresponds to convert the stochastic time evolution of the
density matrix $\rho _{st}(t)$ into the stochastic time evolution of
operators. All this rests on the fundamental relation%
\begin{equation}
\overline{A(t)}=\mathrm{Tr}_{S}[A(0)\rho _{S}(t)]=\mathrm{Tr}_{S}[\rho
_{S}(0)A(t)],  \label{fundamental}
\end{equation}%
where the mean value $\overline{A(t)}$ of a system operator $A$ can be
written in terms of the initial density matrix $\rho _{S}(0)$ and of the
evolved operator $A(t).$

Let us define the dual superoperators $\mathcal{L}_{S}^{\#}$ and $\mathcal{E}%
^{\#}$ by the relations 
\begin{subequations}
\begin{eqnarray}
\mathrm{Tr}_{S}[Ae^{t\mathcal{L}_{S}}\rho ] &=&\mathrm{Tr}_{S}[\rho e^{t%
\mathcal{L}_{S}^{\#}}A], \\
\mathrm{Tr}_{S}[A\mathcal{E}\rho ] &=&\mathrm{Tr}_{S}[\rho \mathcal{E}%
^{\#}A].
\end{eqnarray}%
Eq.~(\ref{Average}) yields the (averaged over realizations) operator time
evolution 
\end{subequations}
\begin{equation}
A(t+\tau )=\sum_{n=0}^{\infty }\int_{0}^{\tau +t}dt^{\prime }\mathcal{W}%
^{\#(n)}(t^{\prime })\mathcal{P}_{0}^{\#}(\tau +t-t^{\prime })A(0).
\label{DualAverageA}
\end{equation}%
Here, $\mathcal{W}^{\#(n)}(z)=[w(z-\mathcal{L}_{S}^{\#})\mathcal{E}%
^{\#}]^{n} $ arises from Eq.~(\ref{Woperator}), and $\mathcal{P}%
_{0}^{\#}(z)=P_{0}(z-\mathcal{L}_{S}^{\#})$ from Eq.~(\ref{PoOperator}).
Note that, as a consequence of the prescription of Eq.~(\ref{fundamental})
the $\mathcal{W}^{\#}$ superoperator applies after the $\mathcal{P}_{0}^{\#}$
one rather than before it as in Eq.~(\ref{Average}).

In conclusion the dynamics of the stochastic operator $A_{st}$ resembles
that of the stochastic density matrix $\rho _{st}.$ It consists, too, of
time intervals with the time evolution driven by $\exp [t\mathcal{L}%
_{S}^{\#}],$ and of others, corresponding to the action of the superoperator 
$\mathcal{E}^{\#}$, where it is driven by disruptive events. Nevertheless,
notice that when $[\mathcal{L}_{S}^{\#},\mathcal{E}^{\#}]\neq 0,$ the time
ordering of the superoperators is reversed as a consequence of turning the
Schr\"{o}dinger's into the Heisenberg's representation.

\subsection{Operator expectation values and correlations}

For the main purpose of working with simplified expressions, in this section
we make all calculations in the interaction representation with respect to $%
\mathcal{L}_{S}^{\#},$ and we assume that%
\begin{equation}
\lbrack \mathcal{L}_{S}^{\#},\mathcal{E}^{\#}]=0,  \label{Conmuting}
\end{equation}%
which in turn in the Schr\"{o}dinger representation yields $[\mathcal{L}_{S},%
\mathcal{E}]=0,$ or equivalently the commutation condition $[\mathcal{L}_{S},%
\mathcal{L}]=0.$ As a consequence of this condition, Eq.~(\ref{DualAverageA}%
) (Eq.~(\ref{Average})) makes the operator (density matrix) stochastic
dynamics, in the interaction representation with respect to $\mathcal{L}%
_{S}^{\#}$ $(\mathcal{L}_{S}),$ only consist of the application of the
superoperator $\mathcal{E}^{\#}$ $(\mathcal{E}).$ Notice that all the
expressions obtained in the previous section, in the interaction
representation, remain valid by setting $\mathcal{L}_{S}\rightarrow 0.$

The operator expectation values, in the interval $(t,t+\tau )$ are written
as $[\overline{\mathbf{A}(\tau +t)}\rightarrow \overline{\mathbf{A}(\tau )}]$%
\begin{subequations}
\label{mean}
\begin{eqnarray}
\overline{\mathbf{A}(\tau )} &=&\mathrm{Tr}_{S}[\rho _{S}(0)\mathbf{A}(\tau
+t)], \\
&=&\mathrm{Tr}_{S}[\rho _{S}(0)C_{\mathrm{I}\mathbf{A}}(\tau ,t).
\end{eqnarray}%
With $\mathbf{A}=(A_{1},A_{2},\cdots )^{\mathrm{T}},$ we denote a vector of
system operators defining a complete basis in the dual (operators) space. $%
\mathrm{I}$ denotes the system identity\ operator. The operator correlations
are written as 
\end{subequations}
\begin{subequations}
\label{correlator}
\begin{eqnarray}
\overline{O(t)\mathbf{A}(t+\tau )} &=&\mathrm{Tr}_{S}[\rho _{S}(0)O(t)%
\mathbf{A}(t+\tau )],  \label{confusa} \\
&=&\mathrm{Tr}_{S}[\rho _{S}(0)C_{O\mathbf{A}}(\tau ,t)].
\end{eqnarray}%
Here, $O$ denotes an arbitrary system operator. The auxiliary operators
function $C_{UV}(\tau ,t),$ acting on arbitrary system operators $U$ and $V,$
are defined by the expression 
\end{subequations}
\begin{equation}
C_{UV}(\tau ,t)\equiv \sum_{m=0}^{\infty }\sum_{n=0}^{\infty }P(\tau ,m;t,n)(%
\mathcal{E}^{\#})^{n}[U(\mathcal{E}^{\#})^{m}[V]].  \label{Cdefinition}
\end{equation}%
$P(\tau ,m;t,n)$ is the probability that $n$ events occur in the interval $%
(0,t)$ and $m$ events in the interval $(t,t+\tau ).$ Under the condition (%
\ref{Conmuting}), in the interaction representation with respect to $%
\mathcal{L}_{S}^{\#},$ Eq.~(\ref{Cdefinition}) follows straightforwardly
from the stochastic dynamics associated to Eq.~(\ref{DualAverageA}).

The set of probabilities $P(\tau ,m;t,n)$ can be written as%
\begin{equation}
P(\tau ,0;t,n)=\int_{0}^{t}dt^{\prime }P_{0}(\tau +t-t^{\prime
})w^{(n)}(t^{\prime }),
\end{equation}%
when $m=0,$ and as%
\begin{eqnarray}
P(\tau ,m;t,n)\! &=&\!\int_{0}^{\tau }\!d\tau ^{\prime }P_{0}(\tau -\tau
^{\prime })\!\int_{0}^{\tau ^{\prime }}\!d\tau ^{\prime \prime
}w^{(m-1)}(\tau ^{\prime }-\tau ^{\prime \prime })  \notag \\
&&\times \int_{0}^{t}\!dt^{\prime }w(\tau ^{\prime \prime }+t-t^{\prime
})w^{(n)}(t^{\prime }),  \label{n and m>1}
\end{eqnarray}%
for $m\geq 1.$ As before, the function $w^{(n)}(\tau )$ is defined by its
Laplace transform $w^{(n)}(u)=[w(u)]^{n}.$ In the Laplace domain, $(\tau
\rightarrow u,$ $t\rightarrow z),$ after using Eq.~(\ref{DoubleLaplace}), we
get%
\begin{equation*}
P(u,0;z,n)=\frac{1-z\tilde{w}(u,z)}{u}P_{0}(z)[w(z)]^{n},
\end{equation*}%
and for $m\geq 1,$%
\begin{equation*}
P(u,m;z,n)=P_{0}(u)[w(u)]^{m-1}z\tilde{w}(u,z)P_{0}(z)[w(z)]^{n},
\end{equation*}%
where $\tilde{w}(u,z)$ is defined by Eq.~(\ref{w(u,z)}).

After some algebra based on Eq.~(\ref{Cdefinition}) we write 
\begin{equation}
C_{UV}(u,z)=\mathcal{G}^{\#}(z)U\mathcal{G}^{\#}(u)[1+z\Delta (u,z)\mathcal{L%
}^{\#}]V.  \label{CUV}
\end{equation}%
The function $\Delta (u,z)$ is defined by Eq.~(\ref{Delta}) and $\mathcal{G}%
^{\#}(u)$ denotes the propagator%
\begin{equation}
\mathcal{G}^{\#}(u)=\frac{1}{u-K(u)\mathcal{L}^{\#}},
\end{equation}%
where $\mathcal{L}^{\#}\equiv \mathcal{E}^{\#}-1$ is the dual superoperator
associated to $\mathcal{L}$, Eq.~(\ref{Lindblad}). The kernel $K(u)$ follows
from Eq.~(\ref{Kernel}). Therefore, taking into account that these
expressions were derived in an interaction representation with respect to $%
\mathcal{L}_{S},$ we obtain that $\mathcal{G}^{\#}(u)$\ is the dual
propagator associated to $\mathcal{G}(u),$ Eq.~(\ref{PropaDensityMatrix}).

From Eq.~(\ref{CUV}), after introducing the density matrix%
\begin{equation}
\rho _{S}(z)=\mathcal{G}(z)\rho _{S}(0),
\end{equation}%
the mean values Eq.~(\ref{mean}) and correlations Eq.~(\ref{correlator}) read%
\begin{eqnarray*}
\overline{\mathbf{A}(\tau )} &\dot{=}&\mathrm{Tr}_{S}\{\rho _{S}(z)\mathcal{G%
}^{\#}(u)[1+z\Delta (u,z)\mathcal{L}^{\#}]\mathbf{A}\}, \\
\overline{O(t)\mathbf{A}(t+\tau )} &\dot{=}&\mathrm{Tr}_{S}\{\rho _{S}(z)O%
\mathcal{G}^{\#}(u)[1+z\Delta (u,z)\mathcal{L}^{\#}]\mathbf{A}\}.
\end{eqnarray*}%
Here, for the sake of shortening the notation, we use the symbol $\dot{=}$
to indicate that the left and right hand side of the equality are written in
the time and Laplace domain, respectively. These equations yield the desired
expressions for operator expectation values and correlations. They can be
straightforwardly written in terms of density matrix propagators as%
\begin{eqnarray}
\overline{\mathbf{A}(\tau )} &\dot{=}&\mathrm{Tr}_{S}\{\mathbf{A}\mathcal{G}%
(u)[1+\mathcal{L}\Delta (u,z)z]\rho _{S}(z)\},  \label{MeanRenewal} \\
\overline{O(t)\mathbf{A}(t+\tau )} &\dot{=}&\mathrm{Tr}_{S}\{\mathbf{A}%
\mathcal{G}(u)[1+\mathcal{L}\Delta (u,z)z]\rho _{S}(z)O\}.\ \ \ \ \ 
\label{CorrelatorRenewal}
\end{eqnarray}%
In the Markov case, i.e., when $K(u)=\gamma ,$ these results recover the
expressions that follows from a microscopic derivation based on a
Born-Markov approximation \cite{carmichael}. Furthermore, by using the same
calculations steps it is possible to demonstrate that%
\begin{equation}
\overline{O(t)\mathbf{A}(t+\tau )\tilde{O}(t)}\dot{=}\mathrm{Tr}_{S}\{%
\mathbf{A}\mathcal{G}(u)[1+z\Delta (u,z)\mathcal{L}]\tilde{O}\rho
_{S}(z)O\}.\ \ \   \label{TresOperatorCorrelacion}
\end{equation}

From Eqs.~(\ref{MeanRenewal}) and (\ref{CorrelatorRenewal}) it is immediate
to realize that expectation values and correlations have the same dynamical
structure, showing that the classical Onsager regression hypothesis can be
extended to this context. We make this fact even clearer by introducing a
preparation at time $t$ [Eq.~(\ref{Preparation})], thereby implying the
transformation $z\rho _{S}(z)\rightarrow \rho _{\Pi }.$ Then, the
preparation can be interpreted as a sudden fluctuation at time $t.$ The
earlier expressions indicate that the operator correlation dynamics depend
on the dynamical decay of this fluctuation.

\subsection{Evolutions}

We can explicitly show that operator expectation values and correlations
have the same dynamical behavior. Here, we obtain the inhomogeneous
equations of motion. Nevertheless, as in Eq.~(\ref{Homogenea}), they can be
rewritten as homogeneous ones. By defining a matrix $\mathbb{M}$ by the
relation%
\begin{equation}
\mathrm{Tr}_{S}[\mathbf{A}\mathcal{L}O]=\mathbb{M}\mathrm{Tr}_{S}[\mathbf{A}%
O],
\end{equation}%
which acts on the indexes of vector $\mathbf{A},$ from Eq.~(\ref{MeanRenewal}%
) it is possible to get the evolution 
\begin{equation}
\frac{d}{d\tau }\overline{\mathbf{A}(\tau )}=\int_{0}^{\tau }dt^{\prime
}K(\tau -\tau ^{\prime })\mathbb{M}\overline{\mathbf{A}(\tau ^{\prime })}%
+\Gamma _{\mathrm{I}\mathbf{A}}(\tau ,t),  \label{OperatorM}
\end{equation}%
while from Eq.~(\ref{CorrelatorRenewal}), for the correlations it follows%
\begin{eqnarray}
\frac{d}{d\tau }\overline{O(t)\mathbf{A}(t+\tau )} &=&\int_{0}^{\tau }d\tau
^{\prime }K(\tau -\tau ^{\prime })\mathbb{M}\overline{O(t)\mathbf{A}(t+\tau
^{\prime })}  \notag \\
&&+\Gamma _{O\mathbf{A}}(\tau ,t).
\end{eqnarray}%
The inhomogeneous terms $\Gamma _{\mathrm{I}\mathbf{A}}(\tau ,t)$ and $%
\Gamma _{O\mathbf{A}}(\tau ,t),$ taking into account the preparation [Eq.~(%
\ref{Preparation})], follow from%
\begin{equation}
\Gamma _{OA}(\tau ,t)=\Delta (\tau ,t)\mathbb{M}\mathrm{Tr}_{S}[O\mathbf{A}%
\rho _{\Pi }].
\end{equation}%
These expressions show that even in the presence of strong non-Markovian
non-stationary effects, the regression hypothesis is still valid \cite%
{CorrelationAge,agingCTRW}.

\subsection{Discussion}

The previous analysis demonstrates that the condition (\ref{Conmuting})
guarantees the fulfillment of the quantum regression hypothesis. This
constraint is satisfied, for example, by a two-level system with Liouvillian 
$\mathcal{L}_{S}[\bullet ]=-i\omega _{A}[\sigma _{z},\bullet ]/2,$ where $%
\sigma _{z}$ is the $z$-Pauli matrix, $\omega _{A}$ its transition
frequency, and $\mathcal{E}$ (or equivalently$\mathcal{L}[\bullet ])$ define
a dispersive or thermal reservoir (see respectively Eqs. (71) and (74) in
Ref. \cite{AClass}). When the two-level system is subjected to an external
field the regression hypothesis may be broken. In fact, when the condition (%
\ref{Conmuting}) is not fulfilled, the regression hypothesis does not hold
true in any case. Nevertheless, by writing 
\begin{equation}
\mathcal{L}_{S}=\mathcal{L}_{0}+\epsilon \mathcal{L}_{1},
\label{ExpansionNOCommutan}
\end{equation}%
where $[\mathcal{L}_{0},\mathcal{E}]=0,$ and $[\mathcal{L}_{1},\mathcal{E}%
]\neq 0,$ it is possible to prove that to first order in the parameter $%
\epsilon ,$ the regression hypothesis, independently of the specific
structure of $\mathcal{L}_{1},$ is still valid. In the Schr\"{o}dinger
representation, the operator expectation of Eq.~(\ref{MeanRenewal}) and the
operator correlation of (\ref{CorrelatorRenewal}), to first order in $%
\epsilon ,$ read%
\begin{eqnarray}
\overline{\mathbf{A}(\tau )} &\dot{\approx}&\mathrm{Tr}_{S}\{\mathbf{A}%
\mathcal{G}(u)[\rho _{S}(z)+\mathrm{I}_{\rho }(u,z)]\},  \label{operador} \\
\overline{O(t)\mathbf{A}(t+\tau )} &\dot{\approx}&\mathrm{Tr}_{S}\{\mathbf{A}%
\mathcal{G}(u)[\rho _{S}(z)+\mathrm{I}_{\rho }(u,z)]O\}.\ \ \ \ \ 
\label{correlacion}
\end{eqnarray}%
Note that $\mathcal{G}(u)$ and $\mathrm{I}_{\rho }(u,z)$ are defined by Eq.~(%
\ref{PropaDensityMatrix}) and (\ref{Inhomogeneous}), respectively. After
performing the preparation at time $t,$ the inhomogeneous term follows from
Eq.~(\ref{InhoPrepa}).

The work of Ref. \cite{JSP-QRT} discussed the validity of the regression
hypothesis in the context of non-Markovian dynamics based on Lindblad rate
equations \cite{LindbladRate}. It was also found (see Sec. 6) that an
external (non-commuting) field breaks its applicability. Nevertheless, in
contrast with the present formalism, the non-Markovian effects admit an
underlying Markovian description. Furthermore, the regression theorem was
studied by analyzing the dynamics at the initial and at the asymptotic time.
In that case, the vanishing of the inhomogeneous term is a necessary
condition for the validity of the regression hypothesis. In spite of these
differences, both formalisms lead to consistent and non-contradictory
results.

We remark that similar conditions but no equivalent to Eq.~(\ref{Conmuting})
were found in different contexts. In Refs. \cite{Oconell,Ford,LaxOpt} the
validity of the quantum regression hypothesis beyond a weak coupling regime
was discussed \cite{JSP-QRT}. In Ref. \cite{alonso}, a commutation property
between the system-bath interaction and the system operators was derived by
using a stochastic wave vector formalism and taking into account a Bosonic
bath described in a rotating wave approximation. All these results suggest
that, beyond a Markovian regime, the validity of the quantum regression
hypothesis strongly may depends on the underlying microscopic dynamic.
Nevertheless, the searching of general applicable criteria should not be
discarded \cite{JSP-QRT}.

\section{Linear response theory}

Here we analyze the response of the system, whose density matrix evolution
is given by Eq.~(\ref{Master}), to an external time dependent perturbation.
In the stationary case \cite{Kubo,KuboBook}, the system response to weak
external perturbations is expressed in terms of response functions that are
proportional to the cross correlation function between the variable of
interest and a system variable coupled to the external field. Here, we show
that a similar result can be established, but that, nevertheless, strong
departures from the predictions of the stationary theory may arise.

To simplify the analysis, in the following calculations we assume that $t=0,$
i.e., that the system-environment coupling (initialization of the renewal
dynamics) coincides with the preparation time, and that the coupling with
the external field is switched on at the same time. In this case the
absolute time coincides with the distance $\tau $ from the
system-environment coupling. The average system state $\rho _{S}(\tau )$\ is
written as a series in the external perturbation%
\begin{equation}
\rho _{S}(\tau )\simeq \rho _{S}^{(0)}(\tau )+\lambda \rho _{S}^{(1)}(\tau
)+\cdots .  \label{Series}
\end{equation}%
The parameter $\lambda $ measures the strength of the external perturbation. 
$\rho _{S}^{(0)}(\tau )$ corresponds to the dynamics in the absence of the
perturbation. The contribution $\rho _{S}^{(1)}(\tau )$ can be obtained from
an average of the perturbed realizations or from the (perturbed) master
equation defining the density matrix evolution, Eq.~(\ref{Master}). In
general, one may assume that the external perturbation affects either the
unitary or the dissipative dynamics. Hereby we analyze both cases.

\subsection{Perturbing the dissipative dynamics}

Here, we consider the case when the external perturbation affects (or is
coupled to) the dissipative dynamics. The two contributions in Eq.~(\ref%
{Series}) are evaluated by averaging the perturbed stochastic realizations.
The zero-th-order contribution reads%
\begin{equation}
\rho _{S}^{(0)}(\tau )=\sum_{n=0}^{\infty }\int_{0}^{\tau }d\tau ^{\prime }%
\mathcal{P}_{0}(\tau -\tau ^{\prime })\mathcal{W}^{(n)}(\tau ^{\prime })\rho
_{S}(0).
\end{equation}%
Notice that this expression follows straightforwardly from Eq.~(\ref{Average}%
). The first-order contribution is determined by an average over all
possible trajectories, in each of which the external perturbation acts only
once, to fit the request of a linear response. Then, $\rho _{S}^{(1)}(\tau )$
becomes the double sum%
\begin{eqnarray}
\rho _{S}^{(1)}(\tau )\! &=&\!\sum_{n=0}^{\infty }\sum_{m=0}^{\infty
}\int_{0}^{\tau }d\tau _{1}\int_{0}^{\tau _{1}}d\tau _{2}\int_{0}^{\tau
_{2}}d\tau _{3}\mathcal{P}_{0}(\tau -\tau _{1})  \notag \\
&&\!\!\times \mathcal{W}^{(n)}(\tau _{1}-\tau _{2})\mathcal{O}(\tau
_{2},\tau _{3})\mathcal{W}^{(m)}(\tau _{3})\rho _{S}(0).\ \ \ \ \ 
\label{SuperO}
\end{eqnarray}%
Each sum takes into account all possible events, preceding and ensuing the
action of the external perturbation. In this expression, $\mathcal{P}%
_{0}(\tau )$ and $\mathcal{W}^{(n)}(\tau )$ are defined by Eqs.~(\ref%
{Woperator}) and (\ref{SurvivalOperator}) respectively. The influence of the
external perturbation is described by the superoperator $\mathcal{O}(\tau
_{2},\tau _{3}).$ This is done either by making the superoperator $\mathcal{E%
}$ change with time, without affecting the times of event occurrence, or by
allowing the external stimulus to slightly change the times of event
occurrence, namely a little bit earlier or later, according to the system's
state.

\subsubsection{Perturbing the event superoperator}

In Ref. \cite{error1} the linear response theory was analyzed on the basis
of a classical two-level system where the perturbation does not affect the
time of occurrence of an event, but that the coin tossing selecting the
fluctuations sign has a time-dependent bias. In Section \ref{section2} we
have seen that the Krauss operators $\mathcal{E}[\rho ]$ signal the
occurrence of renewal events. To realize a perturbation on the system of the
same nature, we have to assume that the time of occurrence of collisional
events is not affected by the external perturbation, but that the specific
form of $\mathcal{E}[\rho ]$ is. Then, the Krauss operator is written as%
\begin{equation}
\mathcal{E}(\tau )[\rho ]=\mathcal{E}[\rho ]+\lambda \mathcal{O}(\tau )[\rho
],  \label{SuperTimeDependent}
\end{equation}%
where the superoperator $\mathcal{O}(\tau )$\ satisfies $\mathrm{Tr}_{S}\{%
\mathcal{O}(t)[\rho ]\}=0.$ For simplicity, we assume%
\begin{equation}
\mathcal{O}(\tau )=\xi (\tau )\mathcal{O}.  \label{Separable}
\end{equation}%
$\xi (\tau )$ is a scalar function that defines the temporal dependence of
the external perturbation. The superoperator $\mathcal{O}(\tau _{2},\tau
_{3})$ appearing in Eq.~(\ref{SuperO}) can then be written as%
\begin{equation}
\mathcal{O}(\tau _{2},\tau _{3})=\mathcal{O}(\tau _{2})\exp [(\tau _{2}-\tau
_{3})\mathcal{L}_{S}]w(\tau _{2}-\tau _{3}).  \label{StandardPerturbation}
\end{equation}%
By working in the Laplace domain on the contributions to each sum of Eq.~(%
\ref{SuperO}), after some algebra, we get%
\begin{eqnarray}
\rho _{S}(\tau ) &\simeq &\mathcal{G}(\tau )\rho _{S}(0)+\lambda
\int_{0}^{\tau }d\tau ^{\prime }\mathcal{G}(\tau -\tau ^{\prime })\mathcal{O}%
(\tau ^{\prime })  \label{PrimerOrder} \\
&&\times \int_{0}^{\tau ^{\prime }}d\tau ^{\prime \prime }K(\tau ^{\prime
}-\tau ^{\prime \prime })e^{(\tau ^{\prime }-\tau ^{\prime \prime })\mathcal{%
L}_{S}}\mathcal{G}(\tau ^{\prime \prime })\rho _{S}(0).  \notag
\end{eqnarray}%
The propagator $\mathcal{G}(\tau )$ is defined in the Laplace domain by Eq.~(%
\ref{PropaDensityMatrix}), while the kernel $K(\tau )$ is defined by Eq.~(%
\ref{Kernel}).

Eq.~(\ref{PrimerOrder}) generates the system's response to first order in
the perturbation strength, $\lambda ,$ and consequently the system's linear
response. For the operator expectation values, after introducing the
assumption (\ref{Separable}), we get%
\begin{equation}
\overline{\mathbf{A}(\tau )}=\overline{\mathbf{A}_{0}(\tau )}+\lambda
\int_{0}^{\tau }d\tau ^{\prime }\chi _{\mathbf{A}\mathcal{O}}(\tau ,\tau
^{\prime })\xi (\tau ^{\prime }),
\end{equation}%
where the zero-th-order contribution reads $\overline{\mathbf{A}_{0}(\tau )}=%
\mathrm{Tr}_{S}[\mathbf{A}\mathcal{G}(\tau )\rho _{S}(0)],$ and the response
function is given by%
\begin{equation}
\chi _{\mathbf{A}\mathcal{O}}(\tau ,\tau ^{\prime })=\mathrm{Tr}_{S}[\mathbf{%
A}\mathcal{G}(\tau -\tau ^{\prime })\mathcal{O}\rho _{f}(\tau ^{\prime })].
\end{equation}%
With $\rho _{f}(\tau ^{\prime }),$ we denote%
\begin{equation}
\rho _{f}(\tau ^{\prime })=\int_{0}^{\tau ^{\prime }}d\tau ^{\prime \prime
}K(\tau ^{\prime }-\tau ^{\prime \prime })e^{(\tau ^{\prime }-\tau ^{\prime
\prime })\mathcal{L}_{S}}\mathcal{G}(\tau ^{\prime \prime })\rho _{S}(0).
\label{RhoEfe}
\end{equation}%
Evidently, the response function $\chi _{\mathbf{A}\mathcal{O}}(\tau ,\tau
^{\prime })$\ has the structure of an operator correlation [see Eq.~(\ref%
{TresOperatorCorrelacion})]. This fact becomes more evident when the initial
density matrix corresponds to the stationary state $\rho _{S}^{\infty }$\ of
the unperturbed evolution, i.e.,%
\begin{equation}
\rho _{S}^{\infty }\equiv \lim_{\tau \rightarrow \infty }\mathcal{G}(\tau
)\rho _{S}(0),  \label{EstacionRho}
\end{equation}%
and the condition%
\begin{equation}
\mathcal{L}_{S}[\rho _{S}^{\infty }]=\rho _{S}^{\infty },  \label{Thermal}
\end{equation}%
is satisfied. Then, the expectation values are written as%
\begin{equation}
\overline{\mathbf{A}(\tau )}=\overline{\mathbf{A}_{\infty }}+\lambda
\int_{0}^{\tau }d\tau ^{\prime }\chi _{\mathbf{A}\mathcal{O}}^{\infty }(\tau
,\tau ^{\prime })\xi (\tau ^{\prime }),  \label{ValorMedio}
\end{equation}%
where $\overline{\mathbf{A}_{\infty }}=\mathrm{Tr}_{S}[\mathbf{A}\rho
_{S}^{\infty }].$ The response function becomes%
\begin{equation}
\chi _{\mathbf{A}\mathcal{O}}^{\infty }(\tau ,\tau ^{\prime })=\mathrm{Tr}%
_{S}[\mathbf{A}\mathcal{G}(\tau -\tau ^{\prime })\mathcal{O}\rho
_{S}^{\infty }]f(\tau ^{\prime },0).  \label{FuncionRespuesta}
\end{equation}%
Here, the function $f(\tau ,0)$ is defined by Eq.~(\ref{Sprinkling}). The
quantum statistical average over $\rho _{S}^{\infty }$ can be read as a
correlation between the operator $\mathbf{A}$ and the superoperator $%
\mathcal{O}$ [see Eq.~(\ref{TresOperatorCorrelacion}) with $z\rho
_{S}(z)\rightarrow \rho _{S}^{\infty }$ and $\Delta (u,z)\rightarrow 0].$

For Markovian dynamics [Eq.~(\ref{Markov})], where $f(\tau ^{\prime
},0)\rightarrow \gamma ,$ the response $\chi _{\mathbf{A}\mathcal{O}%
}^{\infty }(\tau ,\tau ^{\prime })$ depends only on $(\tau -\tau ^{\prime
}). $ This stationary condition \cite{Kubo,KuboBook} is broken in the
non-Markovian case. In fact, the presence of the factor $f(\tau ^{\prime
},0) $ implies that $\chi _{\mathbf{A}\mathcal{O}}^{\infty }(\tau ,\tau
^{\prime })$ depends separately on both $\tau $ and $\tau ^{\prime }.$
Depending of the behavior of $f(\tau ^{\prime },0)$ [see Eqs.~(\ref%
{RocioTwoExpo}) and (\ref{SprinklingFraccionaria})] in the long-time regime
the system may become insensitive to the external perturbation. This effect,
some times called death of linear response \cite{csf}, was found in
classical systems in Refs. \cite{GrigoLinearResponse,SokolovKlafer}. The
present analysis leads us to conclude that the same amazing phenomenon may
be observed in quantum systems.

The previous results also follows from calculations based on the density
matrix evolution, Eq.~(\ref{Master}). On the other hand, we assumed that the
external perturbation is switched on at the initial time. The general case,
i.e., when the renewal dynamics start at time $t=0$ and the external
perturbation is switched on at time $t>0,$ can be discussed following the
same calculations steps. The final expressions involve some extra
contributions. Nevertheless, under the assumption (\ref{ExpansionNOCommutan}%
), to first order in $\epsilon ,$\ Eq.~(\ref{PrimerOrder}) remains valid
under the replacements $\rho _{S}(0)\rightarrow \rho _{S}(t),$ and $K(\tau
)\rightarrow K_{t}(\tau ),$ where $K_{t}(u)=u\tilde{w}(u,t)/[1-w(u)].$
Similarly, Eq.~(\ref{FuncionRespuesta}) remains valid under the replacement $%
f(\tau ^{\prime },0)\rightarrow f(\tau ^{\prime },t).$

As an example we consider a two-level system, with states $\{\left\vert \pm
\right\rangle \},$ whose unitary evolution is defined by $\mathcal{L}%
_{S}[\rho ]=-i\Omega \lbrack \sigma _{x},\rho ]/2,$ where $\sigma _{x}$ is
the $x$-Pauli matrix in the basis $\{\left\vert \pm \right\rangle \}.$ The
perturbed superoperator read%
\begin{equation}
\mathcal{E}(\tau )[\rho ]=\frac{1}{2}\sum_{a,b}\sigma _{ab}\rho \sigma
_{ab}^{\dag }[1-a\lambda \xi (\tau )],  \label{Super}
\end{equation}%
where $\sigma _{ab}\equiv \left\vert a\right\rangle \left\langle
b\right\vert ,$ $(a,b)=\pm ,$ and $\lambda |\xi (\tau )|\leq 1.$ It can be
rewritten as $\mathcal{E}(\tau )[\rho ]=[\mathrm{I}+\lambda \xi (\tau
)\sigma _{z}]/2.$ Then, both the unperturbed and the perturbed terms turn
out to be independent of the initial state $\rho .$ The unperturbed dynamics
correspond to a depolarizing channel \cite{nielsen}, with stationary state $%
\rho _{S}^{\infty }=\mathrm{I}/2.$ The perturbed dynamics modulate the
probability of transitions between the two states \cite{error1}.

From Eq.~(\ref{ValorMedio}), the difference between the upper and lower
populations, i.e., the mean value of the $z$-Pauli matrix $\sigma _{z},$ $%
S_{Z}(\tau )=\mathrm{Tr}_{S}[\rho _{S}(\tau )\sigma _{z}],$\ reads%
\begin{equation}
S_{Z}(\tau )=\lambda \!\int_{0}^{\tau }\!\!d\tau ^{\prime }P_{0}(\tau -\tau
^{\prime })\cos [\Omega (\tau -\tau ^{\prime })]f(\tau ^{\prime },0)\xi
(\tau ^{\prime }).  \label{PromedioZeta}
\end{equation}%
As the transformation $\mathcal{E}(\tau )[\rho ]$ does not depend on $\rho ,$
it is simple to prove that Eq.~(\ref{PromedioZeta}) also corresponds to the
exact solution to all orders in $\lambda .$ On the other hand, notice that
the results obtained in Ref. \cite{error1}, are recovered in the limit $%
\Omega \rightarrow 0.$ Using Eq.~(\ref{Super}), it is easy to establish the
nature of the stochastic dynamics associated to Eq.~(\ref{PromedioZeta}),
i.e., $S_{Z}(\tau )=\langle S_{st}(\tau )\rangle ,$ where $\left\langle
\cdots \right\rangle $ denotes the average over the single realizations.
Between two consecutive events [action of $\mathcal{E}(\tau )$], occurring
at times $\tau _{i-1}$ and $\tau _{i},$ the stochastic evolution is given by 
$S_{st}(\tau )=\cos [\Omega (\tau -\tau _{i-1})]S_{st}(\tau _{i-1}),$ where $%
\tau \in (\tau _{i},\tau _{i-1}),$ while at $\tau =\tau _{i},$ we apply the
disruptive transformation $S_{st}(\tau )\rightarrow \lambda \xi (\tau _{i}).$
The statistics of the time intervals $(\tau _{i}-\tau _{i-1})$ is given by
the distribution $w(\tau ).$ In the result illustrated by the following
figures, we use Eq.~(\ref{WaitFrac}). 
%figura%figura%figura%figura%figurav%figura%figura%figura%figura%figura%figura%figura%figura%figura%figurav%figura%figura%figura%figura%figura
%figura%figura%figura%figura%figurav%figura%figura%figura%figura%figura%figura%figura%figura%figura%figurav%figura%figura%figura%figura%figura
\begin{figure}[tb]
\includegraphics[bb=5 0 845 1240,angle=0,width=7.0 cm]{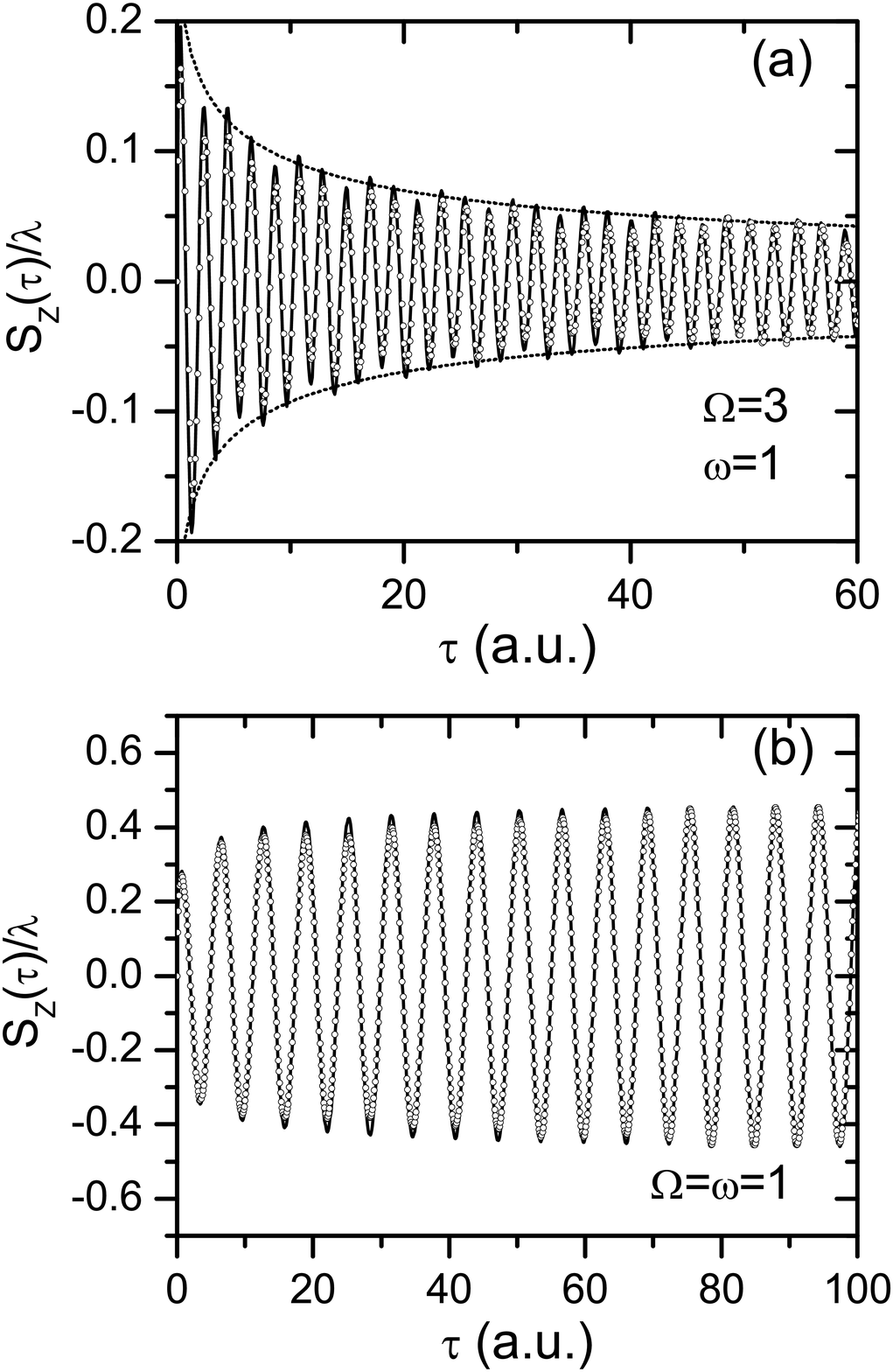}
\caption{Mean value of the z-Pauli matrix (full line), Eq.~(\protect\ref%
{PromedioZeta}), driven by a periodic external perturbation, Eq.~(\protect
\ref{Super}) with $\protect\xi (\protect\tau )=\cos [\protect\omega \protect%
\tau ].$ The parameters of the waiting time distribution, Eq.~(\protect\ref%
{WaitFrac}), are $\protect\alpha =1/2,$ and $A_{\protect\alpha }=1/2.$ The
strength of the perturbation is $\protect\lambda =0.1.$ The circles
correspond to an average over 10$^{3}$ realizations (see text). In (a) the
parameters are $\Omega =3,$ $\protect\omega =1.$ The dotted lines are
proportional to $\pm P_{0}(\protect\tau ).$ In (b), $\Omega =\protect\omega %
=1.$ The time $\protect\tau $ and the parameters are measured in arbitrary
units (a.u.).}
\end{figure}
%figura%figura%figura%figura%figurav%figura%figura%figura%figura%figura%figura%figura%figura%figura%figurav%figura%figura%figura%figura%figura
%figura%figura%figura%figura%figurav%figura%figura%figura%figura%figura%figura%figura%figura%figura%figurav%figura%figura%figura%figura%figura

In Fig.~3 we show both the solution of Eq.~(\ref{PromedioZeta}) and the
average on the realizations of the stochastic simulation. The function $%
f(\tau ,0)$ is given by Eq.~(\ref{SprinklingFraccionaria}). The external
perturbation is $\xi (\tau )=\cos [\omega \tau ].$ We note that, in general,
in the long-time regime the system becomes insensitive to the external
perturbation [Fig.~3a]. In fact, in the Markovian case, or when $\lim_{\tau
\rightarrow }f(\tau ,0)>0,$ the asymptotic behavior of the mean value $%
\lim_{\tau \rightarrow \infty }S_{Z}(\tau )$ is given by an oscillatory
function. In contrast, here $\lim_{\tau \rightarrow \infty }S_{Z}(\tau )=0.$
The decay to this asymptotic value is given by a power law function. The
oscillation amplitude is proportional to the survival probability, Eq.~(\ref%
{PoOperator}), which here can be written as $P_{0}(\tau )=\exp
[A_{1/2}^{2}t] $\textrm{erfc}$[A_{1/2}t^{1/2}].$ In the time asymptotic
regime it behaves as $P_{0}(\tau )\approx 1/(A_{1/2}\sqrt{t})$ \cite{metzler}%
. Only when $\omega =\Omega ,$ the asymptotic behavior is given by an
undamped oscillatory function [Fig.~3b]. The presence of an undamped
asymptotic contribution follows straightforwardly from Eq.~(\ref%
{PromedioZeta}) after expanding the involved trigonometric functions and
using that $P_{0}(u)f(u,0)=1/u-P_{0}(u).$ The non vanishing contribution is $%
[1-P_{0}(\tau )]\cos [\omega \tau ]/2.$ Thus, the convergence to the maximal
amplitude oscillation (one half) also follows a power law behavior. This
effect is seen in Fig.~3b.

\subsubsection{Perturbing the times of events occurrence}

In Refs. \cite{GrigoFluctDiss,GrigoLinearResponse}, the response of a
classical two-level system was analyzed by assuming that the external
perturbation affects the times of event occurrence. The prediction generated
by this assumption has been recently confirmed by experimental results on
liquid crystals \cite{Experiments}. Here, that assumption corresponds to
assuming that the times of the superoperator $\mathcal{E}$'s action are
slightly changed by the external perturbation. Below, we discuss a
system-bath modeling where this condition applies.

We consider a system, which may also have its own (Markovian) dissipative
dynamics, and whose interaction with a complex bath only occurs when the
environment undergoes a structural change, implying the application of $%
\mathcal{E}$ over the system density matrix. The changes between the
different structures of the bath are described by a complex landscape. The
escape over a single well is described by the standard Kramers theory. By
turning on an external perturbation, the height $V$\ of a given well is
written as $V(t)=V_{0}+\lambda _{0}\xi (t).$ By assuming an adiabatic
regime, the survival probability associated to each well evolves as%
\begin{equation}
\frac{d}{d\tau }P_{0}(\tau |t)\simeq -\gamma _{0}(\tau +t)P_{0}(\tau |t).
\label{previousequation}
\end{equation}%
Here, $P_{0}(\tau |t)$ defines the conditional probability that no event
(structural change) occurs in the time interval $(t,t+\tau )$ given that the
last event occurred at time $t.$ The time dependent rate is written as $%
\gamma _{0}(t)=\gamma _{0}\exp [-V(t)/D].$ The coefficient $D$ involves the
temperature and extra parameters describing the well shape. The solution of
Eq.~(\ref{previousequation}) reads $P_{0}(\tau |t)\simeq \exp
[-\int_{0}^{\tau }\gamma _{0}(\tau ^{\prime }+t)d\tau ^{\prime }].$ To first
order in $\lambda _{0},$ it follows%
\begin{equation}
P_{0}(\tau |t)\simeq \exp \Big\{-\gamma \Big[\tau +\lambda \int_{t}^{t+\tau
}\xi (t^{\prime })dt^{\prime }\Big]\Big\}.  \label{PoissonDriven}
\end{equation}%
where $\gamma =\gamma _{0}\exp [-V_{0}/D]$ is the Kramers rate and the
dimensionless strength parameter reads $\lambda =-\lambda _{0}/D.$ If one
assumes a statistical distribution of rate $\gamma $ (due to a random $%
\gamma _{0}$ or $V_{0}$), the survival probability $P_{0}(\tau |t)$ must be
written as a statistical superposition of exponential functions, from which
arbitrary decay behaviors can be recovered. Therefore, Eq.~(\ref%
{PoissonDriven}) can be extended to non-exponential survival probabilities.

In conclusion, the external perturbation shifts the time of event
occurrence, and this property, on the basis of the earlier arguments
corresponds to the assumption%
\begin{equation}
P_{0}(\tau |t)=P_{0}\Big[\tau +\lambda \mathcal{O}\int_{t}^{t+\tau
}dt^{\prime }\xi (t^{\prime })\Big].  \label{Escape}
\end{equation}%
The function\ $\xi (t^{\prime })$ defines the time dependence of the
perturbation. The superoperator $\mathcal{O}$\ takes into account a
dependence of the time shift on the system state.

To first order in $\lambda ,$ the survival probability reads%
\begin{equation}
P_{0}(\tau |t)\approx P_{0}(\tau )-\lambda \mathcal{O}w(\tau
)\int_{t}^{t+\tau }dt^{\prime }\xi (t^{\prime }).  \label{SurvivalPerturbed}
\end{equation}%
Although, in principle, the external perturbation breaks the renewal
character of the process, under the assumption of weak perturbation (small $%
\lambda $), it is legitimate to define the conditional waiting-time
distribution density $w(\tau |t)=-(\partial /\partial \tau )P_{0}(\tau |t),$
thereby getting 
\begin{equation}
w(\tau |t)\approx w(\tau )+\lambda \mathcal{O}\frac{\partial }{\partial \tau 
}\left[ w(\tau )\int_{t}^{t+\tau }dt^{\prime }\xi (t^{\prime })\right] .
\label{WaitPerturbed}
\end{equation}%
Notice that under this approximation, independently of the $\mathcal{O}$
structure, the normalization condition $\int_{0}^{\infty }d\tau w(\tau |t)=1$
is preserved.

The earlier assumptions do not affect Eq.~(\ref{SuperO}), which is made to
remain valid by defining $\mathcal{O}(\tau _{2},\tau _{3})$ as%
\begin{equation}
\mathcal{O}(\tau _{2},\tau _{3})=\mathcal{O}^{\prime }\exp [(\tau _{2}-\tau
_{3})\mathcal{L}_{S}]\delta w(\tau _{2}|\tau _{3}),
\end{equation}%
where $\mathcal{O}^{\prime }=\mathcal{EO}$ and%
\begin{equation}
\delta w(\tau _{2}|\tau _{3})=\frac{\partial }{\partial \tau _{2}}\left[
w(\tau _{2}-\tau _{3})\int_{\tau _{3}}^{\tau _{2}}dt^{\prime }\xi (t^{\prime
})\right] .
\end{equation}%
Nevertheless, in this case, the operator $\mathcal{P}_{0}(\tau -\tau _{1})$
appearing in Eq.~(\ref{SuperO}) also gives a contribution to first order in $%
\lambda ,$ as clearly shown by Eq.~(\ref{SurvivalPerturbed}).

By assuming the initial condition Eq.~(\ref{EstacionRho}), the condition of
Eq.~(\ref{Thermal}), doing the same calculus for the derivation of Eq.~(\ref%
{PrimerOrder}), after rearranging the time integrals and some algebra, the
operator expectation values read%
\begin{equation}
\overline{\mathbf{A}(\tau )}=\overline{\mathbf{A}_{\infty }}+\lambda
\int_{0}^{\tau }d\tau ^{\prime }\chi _{\mathbf{A}\mathcal{O}}^{\infty }(\tau
,\tau ^{\prime })\xi (\tau ^{\prime }),  \label{quantumdynamics}
\end{equation}%
with $\overline{\mathbf{A}_{\infty }}=\mathrm{Tr}_{S}[\mathbf{A}\rho
_{S}^{\infty }]$ and the response function being%
\begin{eqnarray}
\chi _{\mathbf{A}\mathcal{O}}^{\infty }(\tau ,\tau ^{\prime }) &=&\tilde{w}%
(\tau -\tau ^{\prime },\tau ^{\prime })\mathrm{Tr}_{S}\{\mathbf{A}[\mathcal{E%
},\mathcal{O}]\rho _{S}^{\infty }\}  \label{RespuestaPozo} \\
&&-\int_{\tau ^{\prime }}^{\tau }d\tau ^{\prime \prime }\tilde{w}(\tau
^{\prime \prime }-\tau ^{\prime },\tau ^{\prime })  \notag \\
&&\frac{\partial }{\partial \tau ^{\prime \prime }}\left\{ \mathrm{Tr}_{S}[%
\mathbf{A}\mathcal{G}(\tau -\tau ^{\prime \prime })\mathcal{O}^{\prime }\rho
_{S}^{\infty }]\right\} ,  \notag
\end{eqnarray}%
where $\tilde{w}(\tau -\tau ^{\prime },\tau ^{\prime })$ is defined by Eq.~(%
\ref{AgedWait}). Thus, in this case the response function is also
proportional to the correlation between the operator $\mathbf{A}$ and the
external perturbation $\mathcal{O}$.

When $\mathcal{O}^{\prime }\rho _{S}^{\infty }=\rho _{S}^{\infty },$ or $%
\mathcal{O}^{\prime }\rho _{S}^{\infty }=0,$ the integral contribution to
Eq.~(\ref{RespuestaPozo}) vanishes. Then Eq.~(\ref{quantumdynamics}) yields%
\begin{equation}
\overline{\mathbf{A}(\tau )}=\overline{\mathbf{A}_{\infty }}+\lambda 
\overline{\mathbf{A}_{\infty }^{\prime }}\int_{0}^{\tau }d\tau ^{\prime }%
\tilde{w}(\tau -\tau ^{\prime },\tau ^{\prime })\xi (\tau ^{\prime }),
\label{ResponseDynamicoTLS}
\end{equation}%
where $\overline{\mathbf{A}_{\infty }^{\prime }},=\mathrm{Tr}_{S}\{\mathbf{A}%
[\mathcal{E},\mathcal{O}]\rho _{S}^{\infty }\}.$ With the earlier arguments
in mind, we state that this kind of response function is generated whenever
the underlying dynamics can be modeled as an escape process from a well
through a time dependent barrier, or in general when the prescription Eq.~(%
\ref{Escape}) applies. This makes it possible to use for $w(\tau )$ any
form, and not necessarily, the inverse power law form of Refs. \cite%
{GrigoLinearResponse,GrigoFluctDiss}. The example discussed in these papers
is a symmetrical two-level system, with $\mathcal{E}[\rho ]=\sum_{a,b=\pm
}\sigma _{ab}\rho \sigma _{ab}^{\dag }/2=\mathrm{I}/2,$ $\rho _{S}^{\infty }=%
\mathrm{I}/2,$ $\mathcal{O}[\rho ]=-\sum_{a=\pm }a\sigma _{aa}\rho \sigma
_{aa}^{\dag },$ $\mathcal{L}_{S}=0,$ and $\mathbf{A\rightarrow }\sigma _{z}.$
In this case, $\chi _{\mathbf{A}\mathcal{O}}^{\infty }(\tau ,\tau ^{\prime
}) $ follows straightforwardly from the first order contribution associated
to Eq.~(\ref{SurvivalPerturbed}). On the other hand, while Eqs.~(\ref%
{FuncionRespuesta}) and (\ref{RespuestaPozo}) define the system response in
terms of operator correlations, they do not involve in general the
derivative of an operator correlation \cite{EuroPhys}.

\subsection{Perturbing the unitary dynamics}

Now we consider the case where the external perturbation affects the unitary
dynamics acting in the time interval between the occurrence of two
consecutive events. Then, we write%
\begin{equation}
\mathcal{L}_{S}(\tau )=\mathcal{L}_{S}+\lambda \mathcal{L}_{\mathrm{ext}%
}(\tau ).  \label{Unitary}
\end{equation}%
In contrast to the previous case, here we show that when the dynamics
strongly depart from the Markovian case, it is not possible to generate a
linear response theory.

The perturbed dynamics, Eq.~(\ref{Series}), are derived from the master
equation defining the density matrix evolution, Eq.~(\ref{Master}) with $%
t=0. $ To first order in $\lambda ,$ we get 
\begin{equation}
\frac{d\rho _{S}(\tau )}{d\tau }\!\simeq \!\mathcal{L}_{S}(\tau )\rho
_{S}(\tau )+\int_{0}^{\tau }\!\!d\tau ^{\prime }K(\tau -\tau ^{\prime })%
\mathcal{LU}(\tau ,\tau ^{\prime })\rho _{S}(\tau ^{\prime }).  \notag
\end{equation}%
Here, $\mathcal{U}(\tau ^{\prime },\tau ^{\prime \prime })$ is propagator
associated to the time dependent Liouvillian superoperator Eq.~(\ref{Unitary}%
), i.e., $\mathcal{U}(\tau ^{\prime },\tau ^{\prime \prime })=\exp [\mathcal{%
L}_{S}(\tau ^{\prime }-\tau ^{\prime \prime })]+\lambda \mathcal{U}%
^{(1)}(\tau ^{\prime },\tau ^{\prime \prime })+\cdots ,$ where $\mathcal{U}%
^{(1)}(\tau ^{\prime },\tau ^{\prime \prime })$ is the first order
contribution.

The first-order contribution in Eq.~(\ref{Series}), by assuming the
stationary initial condition Eq.~(\ref{EstacionRho}), reads%
\begin{eqnarray}
\rho _{S}^{(1)}(\tau ) &=&\int_{0}^{\tau }d\tau ^{\prime }\mathcal{G}(\tau
-\tau ^{\prime })\Big\{\mathcal{L}_{\mathrm{ext}}(\tau ^{\prime })\rho
_{S}^{\infty }  \label{PerturbedUnitary} \\
&&+\int_{0}^{\tau ^{\prime }}d\tau ^{\prime \prime }K(\tau ^{\prime }-\tau
^{\prime \prime })\mathcal{LU}^{(1)}(\tau ^{\prime },\tau ^{\prime \prime
})\rho _{S}^{\infty }\Big\}.  \notag
\end{eqnarray}%
When calculating the operator expectation values, the first line recovers
the standard Kubo response theory \cite{KuboBook}. On the other hand, the
second line shows that it is impossible to generate a first-order
perturbation. In fact, the validity of this contribution relies on
approximating the difference between the perturbed and unperturbed
propagator, $\mathcal{U}(\tau ^{\prime },\tau ^{\prime \prime })-\exp [%
\mathcal{L}_{S}(\tau ^{\prime }-\tau ^{\prime \prime })],$ by the first
order contribution $\lambda \mathcal{U}^{(1)}(\tau ^{\prime },\tau ^{\prime
\prime }).$ Nevertheless, if the kernel $K(\tau ^{\prime }-\tau ^{\prime
\prime })$\ correlates distant times ($\tau ^{\prime }$ and $\tau ^{\prime
\prime }$), evidently $\mathcal{U}(\tau ^{\prime },\tau ^{\prime \prime })$
cannot be approximated to first order in $\lambda .$ Only when the
non-Markovian dynamics slightly depart from the Markovian condition $[K(\tau
^{\prime }-\tau ^{\prime \prime })=\gamma \delta (\tau ^{\prime }-\tau
^{\prime \prime })],$ the perturbed dynamics can be approximated to first
order in the perturbation. The same conclusion follows by analyzing the
perturbed stochastic trajectories.

\section{Summary and conclusions}

In this paper we have shown that non-standard non-stationary statistical
effects can arise in the context of CP open quantum system dynamics. The
results rely on modelling the system dynamics through a renewal approach,
where the density matrix follows after averaging a set of realizations which
mimic the interaction with a non-Markovian environment. The realizations are
characterized by disruptive abrupt events, producing changes described by
the application of a CP superoperator. The time distance between the
occurrence of two consecutive collisional events is drawn from a non-Poisson
waiting-time distribution density $w(\tau ).$ In the time intervals between
two consecutive collisional events, the system's time evolution is described
by a unitary prescription. Both the CP superoperator and the waiting-time
distribution density take into account the interaction of the system with
the environment.

As a significant advance compared to the earlier work, here we analyzed the
non-Markovian system dynamics by introducing a system preparation at an
arbitrary time and studied the ensuing evolution. The preparation erases the
dependence of the evolution on the previous history of the system.
Nevertheless, it does not erase the memory of the universe, i.e., the
system-environment arrangement. In fact, the master evolution after
preparation depends explicitly on the time preparation [Eqs.~(\ref{Master})
and (\ref{InhoPrepa}) or Eq.~(\ref{Homogenea})]. When the preparation time
is done at arbitrary long times, the ensuing density matrix evolution may or
not converge to an asymptotic structure, the last situation defining the
non-stationary case. It arises when the average time between events is
divergent. When there exists an asymptotic stationary evolution, we showed
that it may significantly depart from the evolution ensuing preparation at
the initial time, i.e., in general the stationary evolution may develop
stronger or weaker non-Markovian effects than the evolution ensuing the
preparation at the initial time.

The possibility of extending the regression hypothesis to the evolutions
arising from the renewal approach was also explored. We showed that when the
unitary dynamics commutes with the event superoperator, the non-Markovian
evolution of expectation and correlation operators are exactly the same
[Eqs.~(\ref{MeanRenewal}) and (\ref{CorrelatorRenewal})]. This result is
valid even in the presence of non-stationary effects. When the commutation
condition is not satisfied, the regression hypothesis remains valid [Eqs.~(%
\ref{operador}) and (\ref{correlacion})] up to first order in the
perturbation [Eq.~(\ref{ExpansionNOCommutan})].

The non-stationary character of the evolution was also analyzed through the
response of the system to an external weak\ perturbation. When the external
field modifies the dissipative dynamics, the response function associated to
the mean value of a given operator can be written as a function of the
correlation between the operator and the external perturbation. Different
response functions [Eqs.~(\ref{FuncionRespuesta}) and (\ref{RespuestaPozo})]
are generated depending on whether we make the perturbation modify the
superoperator structure [Eq.~(\ref{SuperTimeDependent})] without affecting
the occurrence time of the collisional events or we make perturbation affect
the time occurrence of the disruptive events [Eq.~(\ref{Escape})]. As in the
classical counterparts, we have shown that in the presence of non-stationary
dynamics, the response of the system may die out in the time asymptotic
regime. We also concluded that when the external perturbation modifies the
unitary dynamics between events, the linear response theory is incompatible
with the presence of strong memory effects.

The equations that express the previous results are\ also valid for
classical systems. In fact, all quantum properties disappear if one
disregard the unitary contributions, consider diagonal density matrixes, and
take superoperators that do not break that condition. With respect to
previous analysis \cite%
{CorrelationAge,agingCTRW,GrigoLinearResponse,GrigoFluctDiss,EuroPhys,error1,Grigolini}%
, the present results do not rely on a specific form of the waiting-time
distribution (like the inverse power law forms) neither rely on a classical
two-level system modeling. Thus, the formalism applies even when the events
are defined by differential (Fokker-Planck) operators.

The results found in this paper may have direct experimental implications.
In fact, the renewal dynamics arise trivially in the context of
(non-Markovian) quantum kicked systems \cite{Schomerus}. The main
conclusions arrived at with our analysis may also apply to quantum systems
coupled to complex reservoirs generating decay behaviors without a
characteristic time scale \cite{LindbladRate,SingleMolecule}. While a full
quantum microscopic derivation of the present results is an open problem,
our analysis demonstrates that a reach kind of behaviors may arise when
dealing with non-Markovian non-stationary quantum evolutions. Our
contribution is a consistent attempt to model those issues in the context of
CP open quantum system dynamics.

\section*{Acknowledgments}

The authors thank Welch foundation for financial support of this work
through Grant No. B-1577. A.A.B. also thanks support from CONICET, Argentina.


\begin{thebibliography}{99}
\bibitem{exact} For the distribution Eq.~(\ref{WaitFrac}), the function Eq.~(%
\ref{SolutionWithSurvival}) reads $\tilde{P}_{0}(\tau ,t)=\sum_{k=0}^{\infty
}\frac{[-A_{\alpha }(\tau +t)^{\alpha }]^{k}}{\Gamma (\alpha k+1)}%
[1+B_{k}(\tau ,t)],$ with the coefficients $B_{k}(\tau ,t)=\frac{A_{\alpha
}(\tau +t)^{\alpha }}{\Gamma (\alpha )}\beta \lbrack \frac{t}{\tau +t}%
;\alpha ,1+k\alpha ].$ $\beta \lbrack x;a,b]$ is the incomplete Beta
function.
\end{thebibliography}
\end{document}